# Investigation of shock–shock interaction in hypersonic rarefied flow using direct simulation Monte Carlo method

Yazhong Jiang (姜亚中) [1, a], Tianju Ma (马天举) [2]

[1] *Hubei Key Laboratory of Theory and Application of Advanced Materials Mechanics, School of Science, Wuhan University of Technology, Wuhan, 430070, China*

[2] *Academy of Aerospace Propulsion Technology, Xi'an, 710100, China*

The majority of studies on shock–shock interactions assume the high-Reynolds-number or inviscid condition for the fluid flows. However, the rarefied flows encountered by the hypersonic vehicles in their high-altitude flights call for investigations coupling the shock–shock interaction and the rarefied-gas effect. An in-house direct simulation Monte Carlo (DSMC) solver is employed to simulate a series of hypersonic air flows over a wedge–cylinder configuration at the freestream Mach number of 10. The DSMC simulations cover 15 different Knudsen numbers $Kn_\infty$, spanning from $6.688\times10^{-3}$ to $6.688\times10^{-1}$. At the lowest Knudsen number, the numerical results are validated by the corresponding wind-tunnel experiment, and demonstrate the features of an Edney type IV shock–shock interaction, including the type IV wave pattern, the supersonic jet impingement, the augmentations of surface shear stress, pressure and heat flux, as well as the shifts in the angular positions of the peak shear stress, pressure and heat flux over the cylinder surface. In ascending order of $Kn_\infty$, the overviews and details of the flow fields with and without shock–shock interaction, as well as the distributions of the shear stress, pressure and heat flux over the cylinder surface are displayed and analyzed. The increase of flow rarefaction continuously alters the flow pattern of the shock–shock interaction, in which the wave system gradually losses the abilities to deflect the streamlines or to concentrate the energy in the flow. As the flow becomes more rarefied, the shock–shock interaction will result in smaller augmentation factors and less angular shifts of the peak aerodynamic/aerothermal loads. At the highest $Kn_\infty$ in this study, the augmentation factors for skin friction and heat flux are found to be less than unity. The existence/vanishment of the supersonic jet and its position relative to the cylinder account for the distribution characteristics of aerodynamic/aerothermal loads over the cylinder surface.

## I. INTRODUCTION

The shock–shock interaction phenomenon exists in both the external and the internal flows related to modern hypersonic vehicles. Such interactions can change the flow features and cause sharp increases in the aerodynamic/aerothermal loads on the structure. Consequently, the flight performance and safety of the vehicles can be greatly impacted by the shock–shock interactions. Accurate descriptions and predictions of the shock–shock interactions are crucial for the design of hypersonic vehicles. However, the shock–shock interactions generally produce complex flow patterns, involving shock waves, expansion waves, shear layers, supersonic jets, etc. The interplay of these basic flow processes remains to be further investigated.

---

[a] Electronic mail: jiangyazhong@whut.edu.cn

Early investigations on the shock–shock interaction considered a simple but typical problem, which involves an incident oblique shock generated by a slender body and a bow shock detached from a blunt body. Edney's experimental and theoretical study[1] summarized the phenomena in the interactions of oblique and bow shocks, and proposed a system of classification for various interaction patterns. In particular, the Edney type IV interaction receives continual attention because it gives rise to severe augmentation of the aerodynamic/aerothermal loads over the blunt body surface. Edney's findings and methodology laid the foundation for further studies of shock–shock interactions.

Subsequent studies concern different aspects of the shock–shock interaction. First of all, different kinds of geometrical configurations have been considered. The wedge–cylinder configuration,[2–4] which well matches Edney's framework, has been selected as the geometry in most studies of the shock–shock interaction. Olejniczak et al.[5] studied the shock–shock interaction on the double-wedge geometry using the computational fluid dynamics (CFD) based on the Euler equations. Recently, due to the interest in the inward-turning hypersonic inlet of the airbreathing propulsion system, the three-dimensional (3D) shock interactions over the V-shaped blunt leading edges[6–9] were systematically studied. Olejniczak et al.[5] and Zhang et al.[9] established the classifications of interaction patterns as well as the criteria of pattern transitions for the double-wedge and the V-shaped geometries, respectively. As for more complex geometries, the shock–shock interactions during the stage separation of a two-stage-to-orbit system[10–13] were investigated by experiments and CFD computations. Rich flow phenomena associated with the shock–shock interaction have also been studied. Sanderson et al.[3] discussed the nonequilibrium thermochemistry in the shock–shock interaction under high-enthalpy conditions. The intrinsic unsteadiness[4,6,7] in the shock–shock interactions and the unsteady shock–shock interactions driven by the moving body[10–13] also brought new issues of concern, e.g. the hysteresis[10,11] in the shock–shock interaction caused by the periodically varying geometry. Besides, the modeling of turbulent flows[6,8,9] has also been incorporated into the CFD studies of shock–shock interactions.

The underlying assumption in all above-mentioned works is the continuum hypothesis for describing the gas flows and shock waves. To be specific, conventional studies of shock–shock interactions[14] focus primarily on the high-Reynolds-number regime. Therefore, the theoretical investigations are based on inviscid flow theories (e.g., the Rankine–Hugoniot relations and the three-shock solution), while the CFD computations solve the Euler / Navier–Stokes equations. However, the advancement of aerospace technology arouses increasing interests in the high-altitude applications of hypersonic vehicles. The hypersonic rarefied flow[15,16] in these applications has features that are shared by all hypersonic flows (e.g. shock waves and their interactions), and additionally, it is characterized by the flow rarefaction due to the low density of the atmospheric environment. In rarefied flows, the translational motion of gas molecules can be highly nonequilibrium, leading to the breakdown of Euler / Navier–Stokes equations and their resulting inviscid / viscous gas-dynamic theories.[17,18] Apparently, the rarefied-gas effect can modify the knowledge about shock waves as well as their propagation and interaction.[19] The experimental measurements by Alsmeyer[20] and Pham-Van-Diep et al.[21] revealed the interior phenomenon of a shock wave, and confirmed the validity and necessity of using the kinetic approaches for resolving the shock structure. Hornung[22] pointed out the rarefied-gas effect on the shock reflections, in the sense that the thickness of a shock wave is proportional to the mean free path of gas molecules. Ivanov and his collaborators[23–26] conducted detailed simulations of shock reflections, which fully resolved the mean-free-path scales. A finite-size and essentially 2D zone (named non-Rankine–Hugoniot zone[26]) is identified in the immediate vicinity of the triple point in the Mach reflection, where the flow properties deviate from the predictions of the inviscid three-shock theory. Chen et al.[27] and Qiu et al.[28] also studied the kinetic nonequilibrium effects on the Mach and regular reflections of shock waves, using mesoscopic numerical methods. Numerical simulations by Cao et al.[29] investigated the rarefied-gas effect on the diffraction of shock waves. Among the extensive studies on the hypersonic flow over the double-cone/double-wedge geometry, Tumuklu et al.[30] and Sawant et al.[31] employed the direction simulation Monte Carlo (DSMC) method to accurately resolve the complex shock–shock interactions and shock-wave–boundary-layer interactions. The rarefied-gas effects on the wall (e.g., velocity slip and temperature jump) can also be properly handled using such a kinetic approach. Recently, Liu et al.[32] derived a new set of governing equations from gas-kinetic theory to simulate the shock-wave–boundary-layer interaction, with special attentions paid to the wall boundary condition and the nonequilibrium heat conduction in the slightly rarefied gas.

In practical hypersonic aerodynamic problems, the significance of rarefied-gas effects on the shock–shock interactions can be measured by the Knudsen number $Kn_\infty = \lambda_\infty / L$, which is the ratio of the mean free path of the freestream molecules $\lambda_\infty$ to the characteristic length $L$ of the flow. Shock–shock interactions at non-negligible Knudsen numbers have been encountered in wind-tunnel experiements.[33–36] Pot et al.[33] adopted a wedge–cylinder configuration to experimentally study the Edney-type interactions in the Office National d'Etudes et de Recherches Aérospatiales (ONERA) R5Ch wind tunnel. The flow Knudsen number based on the cylinder diameter is $Kn_\infty = 6.688 \times 10^{-3}$. Riabov and Botin[34] carried out wind-tunnel experiments with similar geometries at a lower density corresponding to $Kn_\infty = 5 \times 10^{-2}$. The patterns of shock–shock interactions under this condition were observed to deviate from the description in the continuum regime. Recently, Cardona et al.[35–36] developed the experimental facility and visualization technologies to study a variety of shock–shock interactions between space debris during their atmospheric reentry. The freestream and geometrical conditions in the experiments corresponded to a Knudsen number $Kn_\infty = 1.4 \times 10^{-2}$, which is in the slip regime. A lot of differences between rarefied and continuum shock–shock interactions were demonstrated, but the physics that explains these differences remained to be deepen.[36]

In addition to these experiments, numerical investigations of the shock–shock interactions in hypersonic rarefied flows using the DSMC or other gas-kinetic based methods have been carried out. Early attempts to simulate the Edney type IV interaction with DSMC include the work by Carlson and Wilmoth.[37] In the DSMC computations by Riabov and Botin,[34] the interactions over a wedge–cylinder configuration were investigated at three different Knudsen numbers: 0.006, 0.012 and 0.048. Under the condition of $Kn_\infty$ = 0.048, the pattern of Edney type IV interaction was not observed for any geometrical parameters. Considering the wealth of data measured in the ONERA experiment,[33] it is desirable to reproduce these results with a DSMC simulation. Moss et al.[38] performed a detailed DSMC simulation in accordance with the conditions in the ONERA experiment ($Kn_\infty$ = 6.688×10$^{-3}$). The comparison of DSMC and experimental results yielded excellent agreement. The features of Edney type IV interaction were clearly identified in this near-continuum case. However, Glass[39] reported a comparative study between the Navier–Stokes solution, the DSMC solution and the experimental data for the ONERA case. It was found that a purely continuum based CFD analysis would lead to remarkable overpredictions of the shock standoff distance and the surface aerodynamic loads. Colonia et al.[40] also computed the Edney type IV interaction at the conditions in the ONERA experiment employing the Rykov gas-kinetic scheme. Besides, the dependency of the shock–shock interaction on the degree of rarefaction was preliminarily explored.

Recently, White and Kontis,[41] and Agir et al.,[42] conducted a series of DSMC simulations to reproduce the results of the ONERA experiment and further investigating the effect of Knudsen number on the phenomena of shock–shock interactions. For this purpose, three different values of $Kn_\infty$ were considered. In addition to experimental condition $Kn_\infty$ = 6.688×10$^{-3}$, the double and four times of the experimental $Kn_\infty$ (i.e., 1.334×10$^{-2}$ and 2.668×10$^{-2}$) had also been simulated. For each Knudsen number, DSMC simulations were performed for eleven different geometrical configurations, by changing the position of the wedge relative to the cylinder. Thus, for each rarefaction level, all possible interaction patterns had been covered. An important observation by Agir et al.[42] is the unsteadiness of shock–shock interactions in the rarefied-flow regime. In this investigation, the unsteady shock–shock interactions took place when the vertical distance between the wedge and the cylinder was reduced to 49 mm or less, at $Kn_\infty$ of 6.688×10$^{-3}$ and 1.334×10$^{-2}$. The mechanism of the unsteady evolution in the wave pattern was revealed by analyzing the time-dependent DSMC data of the flow field. Nevertheless, the increase of Knudsen number made the unsteady flows back to steady states.

Although the coupling of the rarefied-gas effect and the shock–shock interaction has been studied in the foregoing research works,[33–42] some shortcomings of these investigations can be noticed. First, the range of the flow Knudsen number is not wide enough, only corresponding to the slip regime[43] in rarefied gas dynamics. Moreover, the number of different rarefaction levels considered in each of Refs. 33–42 is no more than three, and such sparse experimental/numerical cases are inadequate to resolve the transition of the shock–shock interaction phenomena from the continuum regime to the rarefied regime, as the Knudsen number gradually increases. In fact, the range of Knudsen number that appeals academic/application interests can be wide, but the laws of variation in the shock–shock interaction with respect to $Kn_\infty$ are still unclear.

The purpose of the present investigation is to obtain a comprehensive description of the shock–shock interaction phenomenon in hypersonic rarefied flows over a wedge–cylinder configuration, with detailed numerical results produced by the DSMC method as well as the mechanisms that explain the observations. The freestream Mach number and the geometrical condition are fixed, so that the investigation can focus on the only influencing factor $Kn_\infty$, which varies in a wide range from $6.688 \times 10^{-3}$ to $6.688 \times 10^{-1}$. Fifteen different values of $Kn_\infty$ are considered to ensure the resolution of data. At each $Kn_\infty$, both the undisturbed-cylinder simulation and the wedge–cylinder simulation are performed, and their results are presented in a comparative manner, in order to highlight the augmentation of aerodynamic/aerothermal loads caused by the interaction. The remainder of this paper contains the descriptions of the DSMC method (Sec. II) and the problem setup (Sec. III). Then, Sec. IV shows the simulation results of the ONERA case, followed by the parametric study on the Knudsen number. Mechanisms that explain how rarefaction affects the shock–shock interaction are elaborated. Finally, conclusions are given in Sec. V.

## II. NUMERICAL METHOD

The DSMC method[43–45] is a particle-based numerical approach to reproducing the underlying physics of the Boltzmann equation. As a kinetic equation for a dilute, single-species and monatomic gas, the Boltzmann equation describes the evolution of the velocity distribution function $f(t, \boldsymbol{x}, \boldsymbol{c})$ at time $t$, position $\boldsymbol{x}$, and molecular velocity $\boldsymbol{c}$. In the absence of external force field, it has an integro-differential form of

$$\frac{\partial f}{\partial t} + \boldsymbol{c} \cdot \frac{\partial f}{\partial \boldsymbol{x}} = \int_{\mathbb{R}^3} \int_0^{4\pi} \left[ f(\boldsymbol{\zeta}^*) f(\boldsymbol{c}^*) - f(\boldsymbol{\zeta}) f(\boldsymbol{c}) \right] c_r \sigma \mathrm{d}\Omega \, \mathrm{d}\boldsymbol{\zeta} , \tag{1}$$

where $\boldsymbol{c}$ and $\boldsymbol{\zeta}$ are the pre-collision velocities of two colliding molecules, $c_r$ denotes their relative speed $|\boldsymbol{c} - \boldsymbol{\zeta}|$, $\boldsymbol{c}^*$ and $\boldsymbol{\zeta}^*$ are the post-collision velocities, $\Omega$ represents the solid angle, and $\sigma \mathrm{d}\Omega$ is the differential collision cross section.

Instead of discretizing Eq. (1) and seeking for its deterministic solution, the DSMC method employs computational particles to simulate the Boltzmann equation with Monte Carlo techniques. In fact, each DSMC particle is a representative molecule sampled from a large number (denoted by $F_{\mathrm{num}}$) of real molecules that constitute the gas. The evolution of particles within each time step $[t, t+\Delta t]$ is implemented by decoupling the free moves and the binary collisions of gas molecules. First, the particles move freely without changes in their velocities and internal states. Meanwhile, boundary treatments must be performed once the particles hit the boundary of the computational domain. Then, binary collisions during this time interval are simulated, with a proper collision scheme to randomly select collision pairs within each computational grid cell. Based on the microscopic information carried by particles in each cell, the flowfield properties (e.g., macroscopic velocity, temperature, and local Mach number) can be calculated by the statistical procedure in gas-kinetic theory. Similarly, the distributions of aerodynamic quantities on the solid wall (e.g., pressure, shear stress, and heat flux) can be obtained by sampling the microscopic fluxes formed by incident and reflected particles. Taking advantages of the particle-based implementation, one can readily extend the DSMC method to simulate multi-species polyatomic gases, provided that the chemical species, the rotational energy, the quantized vibrational level, etc. are recorded for each particle.

An in-house DSMC platform has been developed by the first author and its reliability has been validated in previous studies.[18,46,47] The current code handles arbitrary geometries with body-fitted grids and an efficient particle tracking technique.[18] It provides multiple options for modeling the collisions between gas molecules and the reflections of gas molecules on the solid surfaces. In this paper, the no-time-counter (NTC) scheme[43] is used to recover the correct collision rates for all velocity classes. The relaxation rate of molecular vibration is controlled by the Millikan–White model.[48] For inelastic collisions, the Larsen–Borgnakke redistribution model[49] is employed to exchange energy between translational, rotational, and vibrational modes. The collision cross section of a gas molecule and the scattering law are described by the variable hard sphere (VHS) model.[43] The gas–surface interaction is implemented according to the Maxwellian reflection model.[43]

In the present investigation, emphasis is placed on the wall surface properties, including the wall pressure $p_w$, the wall shear stress $\tau_w$, and the wall heat flux $q_w$. Therefore, the formulas for them in the DSMC method are shown here as

$$p_w = -\boldsymbol{e}_n \cdot \frac{F_{\text{num}}\left[\sum_{\text{in}}(m\boldsymbol{c}) - \sum_{\text{re}}(m\boldsymbol{c})\right]}{\Delta S \Delta t_{\text{sample}}}, \quad (2)$$

$$\tau_w = \boldsymbol{e}_t \cdot \frac{F_{\text{num}}\left[\sum_{\text{in}}(m\boldsymbol{c}) - \sum_{\text{re}}(m\boldsymbol{c})\right]}{\Delta S \Delta t_{\text{sample}}}, \quad (3)$$

$$q_w = \frac{F_{\text{num}}\left[\sum_{\text{in}}(mc^2/2 + \varepsilon_{\text{rot}} + \varepsilon_{\text{vib}}) - \sum_{\text{re}}(mc^2/2 + \varepsilon_{\text{rot}} + \varepsilon_{\text{vib}})\right]}{\Delta S \Delta t_{\text{sample}}}, \quad (4)$$

where $m$, $\varepsilon_{\text{rot}}$, and $\varepsilon_{\text{vib}}$ are the mass, the rotational energy and the vibrational energy of a molecule, respectively. The surface element at a certain position on the wall has an aera denoted by $\Delta S$, a local external normal vector $\boldsymbol{e}_n$, and a tangent vector $\boldsymbol{e}_t$. The symbols $\sum_{\text{in}}$ and $\sum_{\text{re}}$ stand for the summation over incident and reflected particles on the area $\Delta S$ during the sample time $\Delta t_{\text{sample}}$. Note that, for the two-dimensional planar simulation in this paper, the tangent direction of the wall boundary can be defined clearly, and $\tau_w$ can be either positive or negative. Only stationary walls are considered in this study.

## III. PROBLEM DESCRIPTION

This investigation begins with a Mach 10 air flow over a wedge–cylinder configuration, which has been experimentally studied by Pot et al.[33] using the low-density wind tunnel in French ONERA. The purpose of simulating the ONERA case is twofold. First, this flow problem serves as the benchmark for the present DSMC code. Second, this case corresponds to the lowest Knudsen number considered in this paper, and it provides a near-continuum example for the shock interactions.

The geometry is sketched in Fig. 1. A circular cylinder with a diameter of $D = 2R = 16$ mm is the generator of the bow shock wave, and a wedge with a half angle of 20° is the generator of the incident shock wave. The length of the ramp is 50.771 mm. The origin of coordinate system is located at the leading edge of the cylinder, and the tip of wedge is at point (−102 mm, −53 mm). The free stream flows in $x$-direction. In the experiment, the distributions of pressure and heat flux on the windward surface of cylinder were measured, both before and after the wedge was placed upstream of the cylinder. As for the numerical simulation, the curved surface can be modeled with good precision and efficiency by using a body-fitted grid. Besides, the flow domain can be tailored to reduce the computational cost. In Fig. 1, the tailored flow domain is shown as the shaded part.

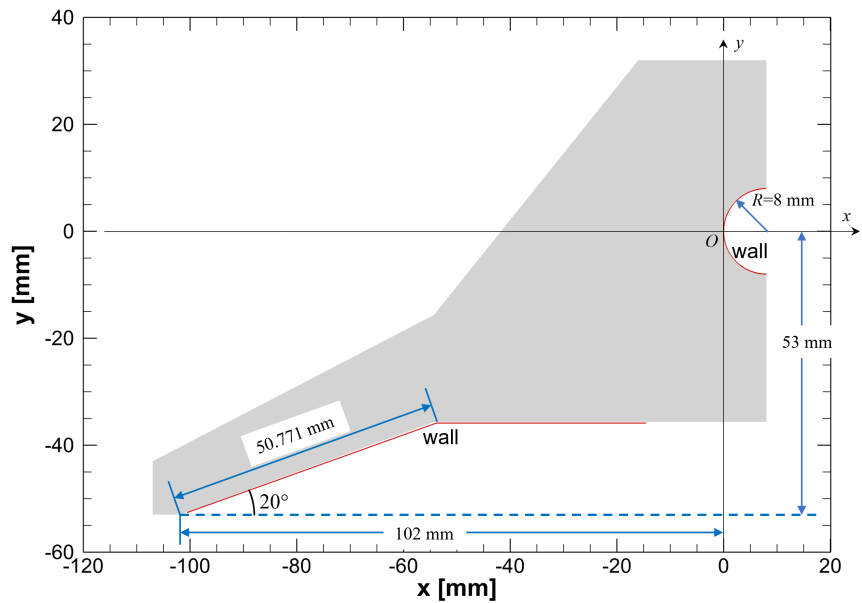

FIG. 1 Geometric configuration, dimensions, coordinate system, and flow domain of the shock–shock interaction problem. The red lines represent the solid wall boundaries.

The working gas in the experiment is air. In the simulation, it is composed of nitrogen and oxygen with molar fractions of 78.8% and 21.2%, respectively. The nitrogen and oxygen molecules are described by the VHS[43] and Millikan–White[48] model with parameters recommended in Refs. 43 and 44. Considering the short duration of the wind-tunnel experiment, all wall surfaces are assumed to be isothermal at room temperature. Diffuse reflection and full thermal accommodation are adopted for gas–surface interactions. The freestream properties and wall conditions are listed in Table I.

TABLE I. Freestream and wall conditions of the baseline case (Case 1) of shock–shock interaction problem.

| Parameter | Value |
| --- | --- |
| Freestream number density $n_\infty$ | $8.140 \times 10^{21}$ m$^{-3}$ |
| Freestream mass density $\rho_\infty$ | $3.899 \times 10^{-4}$ kg/m$^3$ |
| Freestream velocity $u_\infty$ | 1452 m/s |
| Freestream temperature $T_\infty$ | 52.50 K |
| Freestream pressure $p_\infty$ | 5.900 Pa |
| Freestream mean free path $\lambda_\infty$ | 0.1070 mm |
| Freestream mean collision time $\tau_{c,\infty}$ | $5.4 \times 10^{-7}$ s |
| Wall temperature $T_w$ | 300.0 K |
| Diameter of cylinder $D$ | 16.00 mm |
| Freestream Mach number $M_\infty = u_\infty / a_\infty$ | 10 |
| Global Knudsen number $Kn_\infty = \lambda_\infty / D$ | 0.006688 |
| Global Reynolds number $Re_\infty = \rho_\infty u_\infty D / \mu_\infty$ | 1820 |

The dominant control parameters in this investigation are the Mach number $M_\infty$ and the Knudsen number $Kn_\infty$. Another commonly used dimensionless parameter, the Reynolds number $Re_\infty$, is actually determined by the ratio of $M_\infty$ and $Kn_\infty$. Values of these parameters are also listed in Table I, which indicate a hypersonic near-continuum flow feature. Note that the calculation of $Re_\infty$ can be influenced by the viscosity-temperature relation $\mu(T)$ for air. Here, a power-law formula for $\mu(T)$ is used to keep consistent with the VHS[43] model for collision cross sections, and the resulting $Re_\infty = 1820$ is quantitatively different from the value 2658 based on the Sutherland formula.[38]

In addition to the basic simulation in accordance with the ONERA experiment, a series of simulation cases are designed to investigate the flow and aerodynamic characteristics in a wide range of Knudsen number. When the degree of rarefaction varies, the geometric parameters related to the cylinder and wedge remain unchanged, but the freestream density $\rho_\infty$ is altered. Simultaneously, $n_\infty$, $p_\infty$, $\lambda_\infty$, $\tau_{c,\infty}$, $Kn_\infty$, and $Re_\infty$ also vary in direct or inverse proportion to $\rho_\infty$, while other parameters in Table I, including the Mach number, keep the same values. As shown in Table II, fifteen cases are considered, and the magnitude of $Kn_\infty$ varies by two orders. In the fifth column of Table II, the altitudes corresponding to the freestream air densities are provided, according to the data of standard atmosphere.[50] The range of $\rho_\infty$ corresponds to the range of altitude from 58.2 km to 89.3 km. These cases span from the near-continuum regime to the medium-rarefied regime. For each working condition, a pair of simulations were carried out: (a) the flow over a sole cylinder and (b) the wedge–cylinder shock interaction problem. Through the comparison between simulations (a) and (b), the changes in flow pattern and aerodynamic/aerothermal loads caused by the shock–shock interaction can be demonstrated.

TABLE II. Simulation cases covering different degrees of rarefaction. For all cases, the freestream Mach numbers are equal to 10.

| Case | $Kn_\infty$ | $Re_\infty$ | $\rho_\infty$ [kg/m³] | Altitude [km] | Simulations |
|---|---|---|---|---|---|
| 1 | 0.006688 | 1820 | $3.899\times10^{-4}$ | 58.2 | (1a) undisturbed cylinder & (1b) wedge–cylinder interaction |
| 2 | 0.01338 | 910 | $1.949\times10^{-4}$ | 63.6 | (2a) undisturbed cylinder & (2b) wedge–cylinder interaction |
| 3 | 0.01500 | 811 | $1.738\times10^{-4}$ | 64.5 | (3a) undisturbed cylinder & (3b) wedge–cylinder interaction |
| 4 | 0.01800 | 676 | $1.449\times10^{-4}$ | 65.9 | (4a) undisturbed cylinder & (4b) wedge–cylinder interaction |
| 5 | 0.02100 | 580 | $1.242\times10^{-4}$ | 67.0 | (5a) undisturbed cylinder & (5b) wedge–cylinder interaction |
| 6 | 0.02400 | 507 | $1.087\times10^{-4}$ | 68.0 | (6a) undisturbed cylinder & (6b) wedge–cylinder interaction |
| 7 | 0.02675 | 455 | $9.748\times10^{-5}$ | 68.8 | (7a) undisturbed cylinder & (7b) wedge–cylinder interaction |
| 8 | 0.03290 | 370 | $7.926\times10^{-5}$ | 70.3 | (8a) undisturbed cylinder & (8b) wedge–cylinder interaction |
| 9 | 0.04000 | 304 | $6.519\times10^{-5}$ | 71.7 | (9a) undisturbed cylinder & (9b) wedge–cylinder interaction |
| 10 | 0.05169 | 235 | $5.045\times10^{-5}$ | 73.4 | (10a) undisturbed cylinder & (10b) wedge–cylinder interaction |
| 11 | 0.06000 | 203 | $4.346\times10^{-5}$ | 74.4 | (11a) undisturbed cylinder & (11b) wedge–cylinder interaction |
| 12 | 0.06688 | 182 | $3.899\times10^{-5}$ | 75.2 | (12a) undisturbed cylinder & (12b) wedge–cylinder interaction |
| 13 | 0.1500 | 81 | $1.738\times10^{-5}$ | 80.4 | (13a) undisturbed cylinder & (13b) wedge–cylinder interaction |
| 14 | 0.3344 | 36 | $7.798\times10^{-6}$ | 85.3 | (14a) undisturbed cylinder & (14b) wedge–cylinder interaction |
| 15 | 0.6688 | 18 | $3.899\times10^{-6}$ | 89.3 | (15a) undisturbed cylinder & (15b) wedge–cylinder interaction |

Before the section of results and discussion, some assumptions for the simulations of the above flows should be noticed.

First, all flows in this paper are treated as two-dimensional (2D) planar problems. For case 1, which corresponds to the ONERA experiment, the spanwise lengths of the actual wedge and cylinder are both 0.1 m. As a result, the length-to-diameter ratio of the cylinder model is 6.25. In the middle cross section of the cylinder, a 2D approximation is acceptable. Nevertheless, the difference between 2D simulation and 3D experiment will inevitably lead to discrepancies between the numerical and experimental results. Since the 2D approximation is performed throughout the present numerical study, it will not affect the discussion part in this paper.

Second, a steady flow state can be reached in every simulation, so that the time-averaging strategy can be used in the DSMC method to get low-noise macroscopic data. In fact, the DSMC method is inherently time-accurate, so the steadiness of each macroscopic flow field can be confirmed by monitoring the instantaneous output of each DSMC simulation.

Third, for all cases, the total enthalpy of the free stream is only 1.1 MJ/kg, which is much smaller than the dissociation energies of oxygen and nitrogen. Consequently, there is no need to consider chemical reactions herein. Moderate excitation of vibrational energy can be expected, and it is included in the DSMC simulations.

Fourth, for case 1, the Navier–Stokes computation failed to reproduce the experimental results.[39] Therefore, simulations based on kinetic theory are necessary for the flow regimes considered in this paper, even for the lowest $Kn_\infty$.

# IV. RESULTS AND DISCUSSION

Simulations of flow cases in Table II were conducted using the DSMC method described in Section II. According to the guidelines for the DSMC discretization, the computational grid, the time step size, and the number of computational particles were designed properly for each simulation. A summary of the numerical parameters for the shock interaction simulations are shown in Table III, which includes the information about the time step $\Delta t$, the area of the 2D computational domain $A_d$, the total number of grid cells $N_c$, the minimum cell size $\Delta s_{min}$, and the total number of particles $N_p$ at the end of the simulation. With these settings, the kinetic scales in space and time can be fully resolved, and the number of particles per cell is sufficient. As shown in the last column, the ratio of $A_d$ to the cross-section area of the cylinder $\pi R^2$ indicates that the computational domain must be enlarged as the Knudsen number grows, in order to ensure that the flow at the inlet boundary is not disturbed by the shock waves.

TABLE III. The computational parameters in the DSMC simulations of the shock interactions for cases 1–15. For each case, $\Delta t$ and $\Delta s_{min}$ are normalized by the mean collision time $\tau_{c,\infty}$ and the mean free path $\lambda_\infty$ corresponding to the freestream condition, respectively.

| Case | $Kn_\infty$ | $\Delta t/\tau_{c,\infty}$ | $\Delta s_{min}/\lambda_\infty$ | $N_c$ | $N_p$ | $A_d/(\pi R^2)$ |
|---|---|---|---|---|---|---|
| 1 | 0.006688 | $3.683\times10^{-3}$ | $1.869\times10^{-2}$ | 828000 | 48711027 | 20.07 |
| 2 | 0.01338 | $3.682\times10^{-3}$ | $1.869\times10^{-2}$ | 207000 | 13137233 | 20.07 |
| 3 | 0.01500 | $4.106\times10^{-3}$ | $3.125\times10^{-2}$ | 207000 | 18054052 | 26.95 |
| 4 | 0.01800 | $4.106\times10^{-3}$ | $3.125\times10^{-2}$ | 207000 | 18880292 | 26.95 |
| 5 | 0.02100 | $5.865\times10^{-3}$ | $3.572\times10^{-2}$ | 196625 | 11598542 | 26.95 |
| 6 | 0.02400 | $7.698\times10^{-3}$ | $3.907\times10^{-2}$ | 161850 | 8833083 | 32.49 |
| 7 | 0.02675 | $9.209\times10^{-3}$ | $3.505\times10^{-2}$ | 132950 | 7307145 | 32.49 |
| 8 | 0.03290 | $9.359\times10^{-3}$ | $2.850\times10^{-2}$ | 132950 | 7393125 | 32.49 |
| 9 | 0.04000 | $7.698\times10^{-3}$ | $2.344\times10^{-2}$ | 132950 | 7458051 | 32.49 |
| 10 | 0.05169 | $5.957\times10^{-3}$ | $1.814\times10^{-2}$ | 99500 | 5214554 | 38.09 |
| 11 | 0.06000 | $5.132\times10^{-3}$ | $1.563\times10^{-2}$ | 99500 | 5187026 | 38.09 |
| 12 | 0.06688 | $9.208\times10^{-3}$ | $2.336\times10^{-2}$ | 40600 | 5987733 | 57.66 |
| 13 | 0.1500 | $4.106\times10^{-3}$ | $1.042\times10^{-2}$ | 40600 | 6070321 | 57.66 |
| 14 | 0.3344 | $3.683\times10^{-3}$ | $2.804\times10^{-2}$ | 11700 | 615757 | 110.5 |
| 15 | 0.6688 | $1.842\times10^{-3}$ | $1.402\times10^{-2}$ | 11700 | 654249 | 110.5 |

Experimental data[33] for Case 1 were used to validate the numerical methodology, while the reliability of the DSMC results for Cases 2–15 was verified through the parameter independence studies. In this section, only the baseline flow (Case 1) is presented in a complete and detailed manner. The numerical results of the other cases are selected and organized to highlight the effects of increasing rarefaction on the shock–shock interaction.

## A. Baseline case

Figures 2 and 3 depict the flow field obtained by the DSMC simulation for Case 1. Figure 2a shows typical flow phenomena in a hypersonic near-continuum flow over an undisturbed cylinder. A slight discrepancy between the rotational and translational temperatures can be observed, while the vibrational nonequilibrium is remarkable. When the bow shock and the incident oblique shock interact (Fig. 2b), the shock stand-off distance is enlarged significantly, and a complex flow pattern is formed.

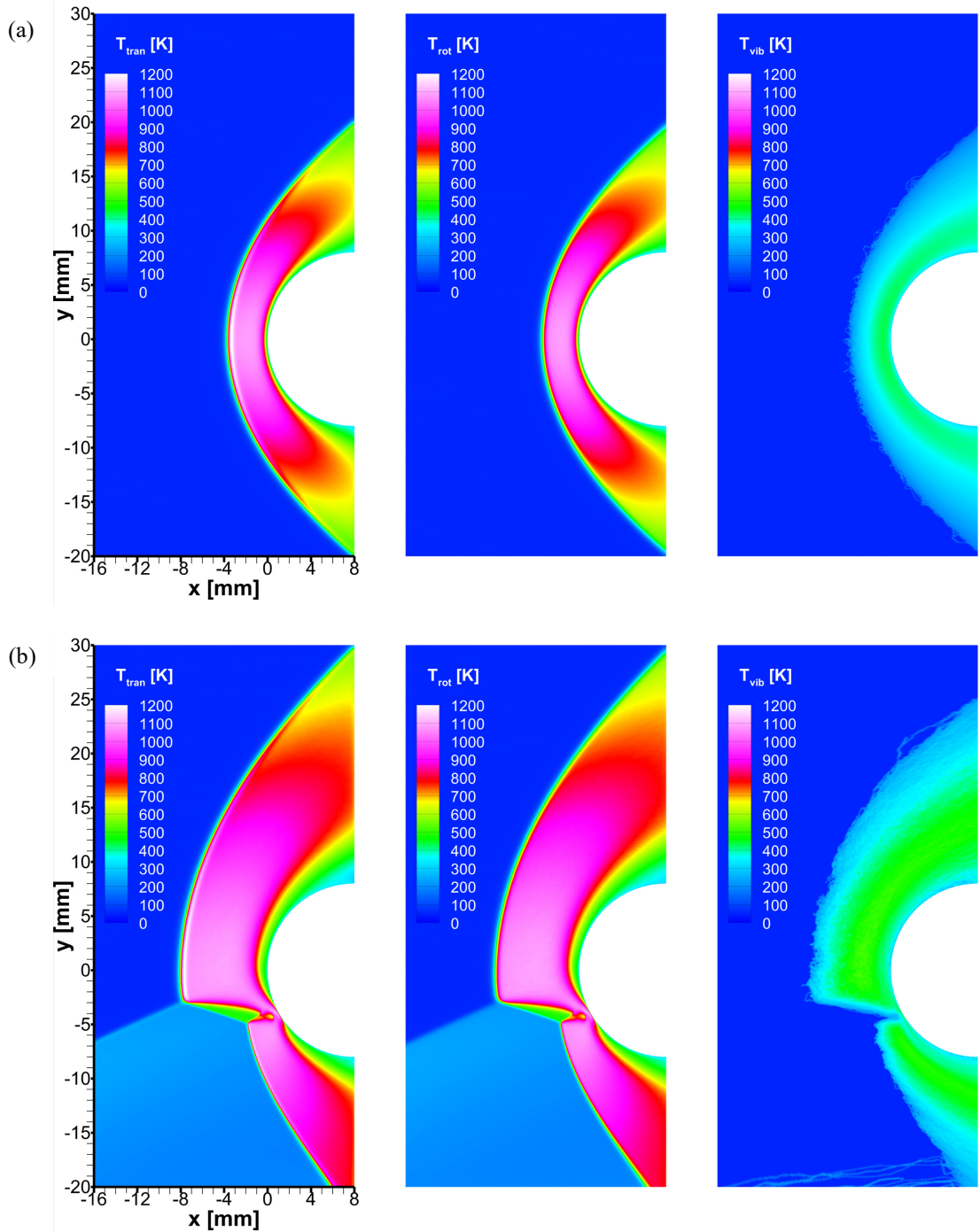

FIG. 2 Translational, rotational, and vibrational temperature fields for Case 1: (a) without incident shock wave and (b) interacting with the incident shock wave.

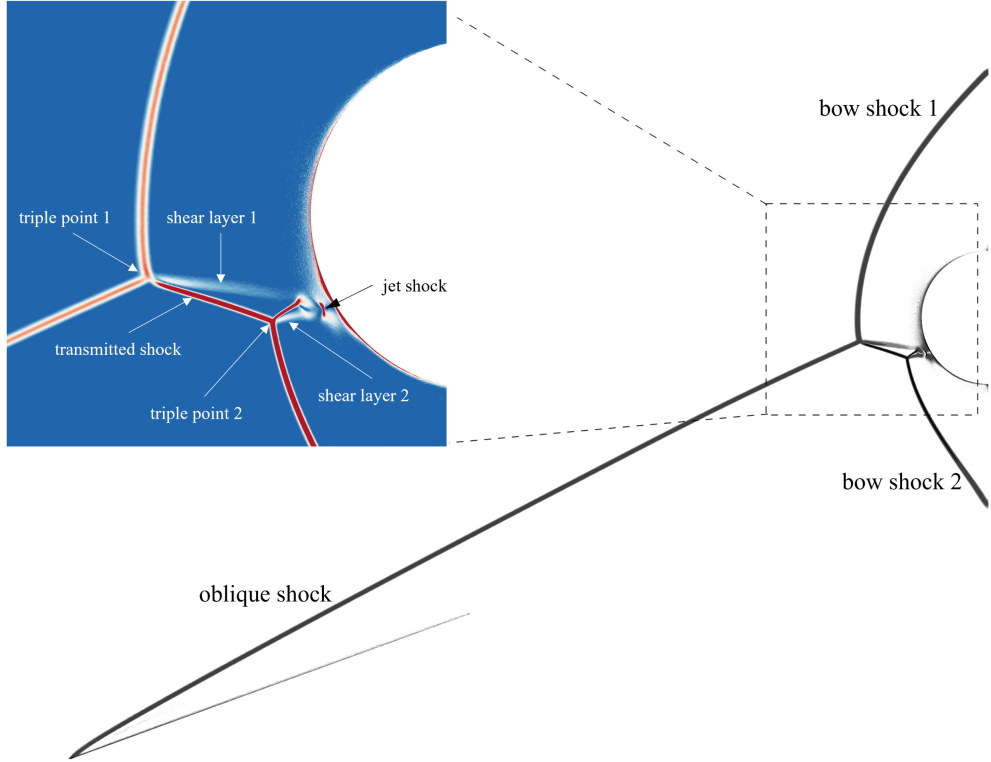

FIG. 3 Numerical schlieren of the shock–shock interaction (Case 1). The inset shows the wave structure in the vicinity of the cylinder.

As shown in Fig. 2(b) and Fig. 3, the structure of the wave system for this near-continuum case can be effectively described by the Edney type IV interaction pattern, although the latter is based on inviscid flow theory. Upon the impingement by the oblique shock, the bow shock is split into two portions, which are connected by a transmitted shock. Two triple-shock structures are formed, and the shear layers 1 and 2 emanate from the triple points 1 and 2, respectively. Between the two shear layers, a supersonic jet rushes at the cylinder surface, leading to the formation of a jet shock in front of the solid wall. In Fig. 3, the shock waves and the shear layers all have visible thickness, although they are thin enough due to the low rarefaction in Case 1.

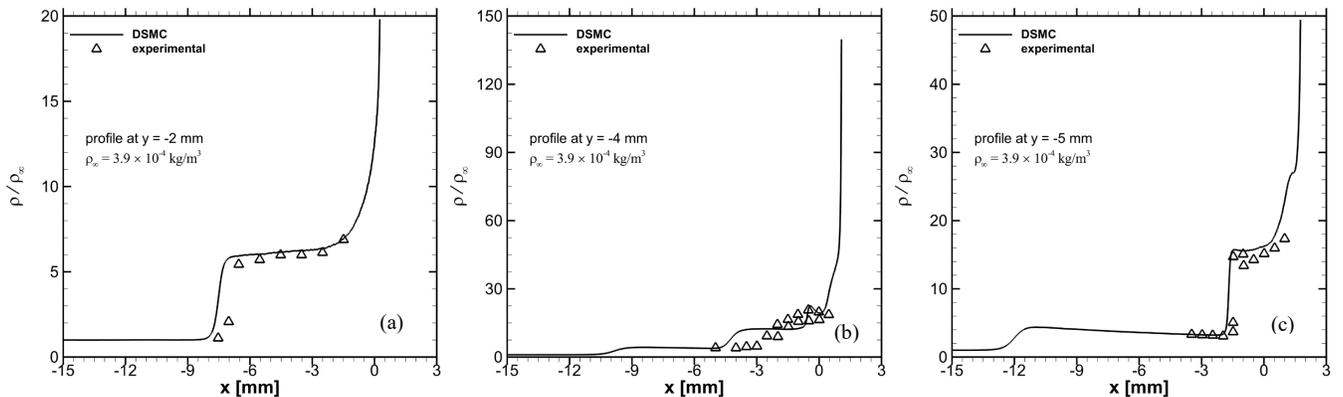

FIG. 4 Normalized density profiles along three different horizontal lines: (a) $y = -2$ mm, (b) $y = -4$ mm, and (c) $y = -5$ mm. The lines $y = -2$ mm and $y = -5$ mm pass through the triple points 1 and 2, respectively. The DSMC results are compared with the experimental data.[33]

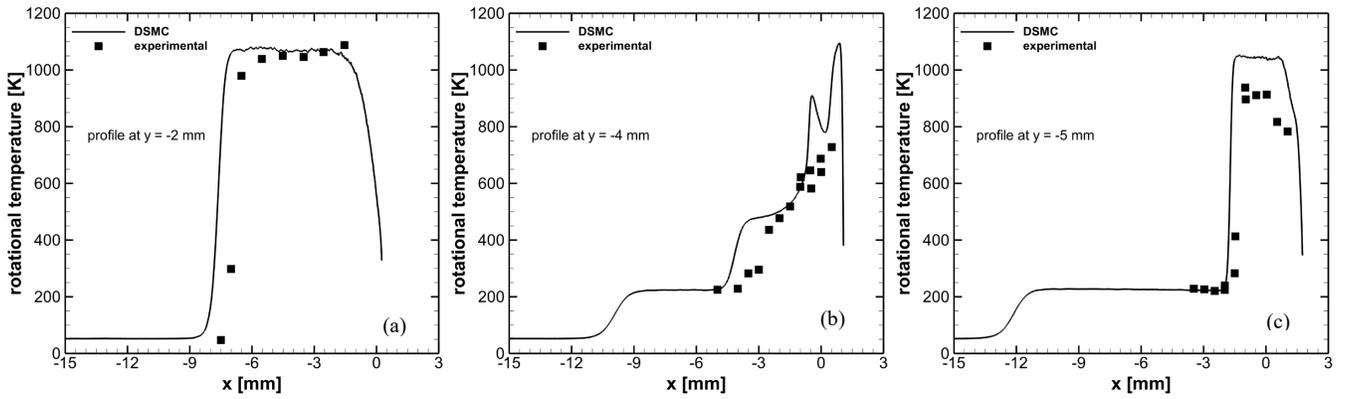

FIG. 5 Rotational temperature profiles along three different horizontal lines: (a) $y = -2$ mm, (b) $y = -4$ mm, and (c) $y = -5$ mm. The lines $y = -2$ mm and $y = -5$ mm pass through the triple points 1 and 2, respectively. The DSMC results are compared with the experimental data.[33]

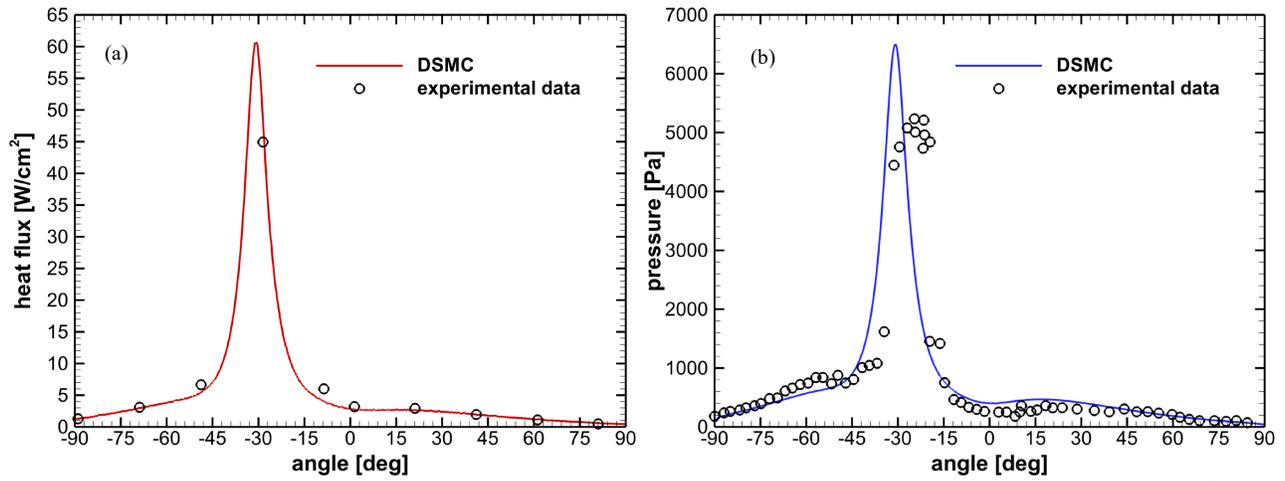

FIG. 6 Distributions of the properties over the cylinder surface: (a) heat flux and (b) pressure. The angle is defined as $\theta = \sin^{-1}(y/R)$. The DSMC results are compared with the experimental data.[33]

In the experiment, gas density $\rho$ and rotational temperature $T_{rot}$ were measured along three specific horizontal lines: $y = -2$ mm, $y = -4$ mm, and $y = -5$ mm. The normalized density data $\rho/\rho_\infty$ along three different lines are plotted in Fig. 4, in comparison with the corresponding DSMC results. Similarly, Figure 5 shows three profiles of $T_{rot}$. Reasonable agreement between the DSMC and experimental data indicates the correctness of the flowfield computation by the DSMC code.

The shock–shock interaction can result in amplified aerodynamic loads and heat transfer rate on the cylinder surface. Figure 6 presents the DSMC results of the heat flux and the pressure distributions along the cylinder surface. The angle $\theta$ is defined so that the lowest point of the cylinder corresponds to $\theta = -90°$. Distributions of the pressure and the heat flux are highly localized, with prominent peak values at about $\theta = -30°$. This can be explained by the supersonic jet impingement on the wall shown in Figs. 2 and 3, which characterized the Edney type IV interaction. The DSMC predictions of the wall pressure and heat flux match the experimental measurements well, and therefore the present numerical method can be regarded as a reliable tool for this investigation.

## B. Effects of increasing Knudsen number

This subsection focuses on how the rarefaction effect influences the shock–shock interaction. The overall flow fields under different Knudsen numbers are demonstrated, followed by detailed flow features in the vicinity of the cylinder. Distributions of aerodynamic force and heating on the cylinder surface are displayed, and their variation characteristics are explained based on the observation of the flowfield structure.

### 1. Flow field

Results of cases 1–15 are illustrated by Fig. 7. For each Knudsen number, the temperature fields are shown to represent the flow structures with and without interaction. As the rarefaction becomes stronger, the shock waves and the shear layers grow thicker and thicker. Meanwhile, the thickness of the boundary layers on the wall surfaces (the cylinder and the wedge) and the shock standoff distance also increase. A quantitatively analysis of the thickness of shock $d$, the thickness of wall boundary layer $\delta$, and the shock standoff distance $\Delta$ indicates that[17]

$$\frac{d}{R} \propto Kn_\infty, \qquad \frac{\delta}{R} \propto \sqrt{Kn_\infty}, \qquad \frac{\Delta - \Delta_{\text{inv}}}{R} \propto \sqrt{Kn_\infty}, \qquad (5)$$

where $\Delta_{\text{inv}}$ denotes the standoff distance under the inviscid condition. Therefore, with the increase of $Kn_\infty$, the interior structures of the shock wave and the boundary layer interact, merge, and eventually occupy the whole region between the wave front and the wall. As can be seen in Figs. 7(m)–(o), the flow fields are actually smooth and highly diffusive. During the transition from Case 1 to Case 15, the geometry and the flow Mach number stay unchanged, but the sole effect of rarefaction can substantially change the pattern of shock–shock interaction. In addition, the gas rarefaction also influences the flow over the wedge. The shape and position of the incident shock wave gradually depart from those predicted by the inviscid oblique-shock theory. Consequently, the intersection point of the incident shock and the bow shock moves upward as $Kn_\infty$ increases.

More details can be observed by inspecting the flow fields near the cylinder. Figure 8 displays the contours of velocity magnitude and the directions of the slip velocity on the cylinder surface. Remarkable velocity slips at the wall boundary can be noticed for all Knudsen numbers considered in this paper. According to the directions of the velocity vectors, the location of the flow stagnation point can be determined. As $Kn_\infty$ increases, the stagnation point firstly moves clockwise, approaching the point at $\theta \approx 0°$, and then it moves counterclockwise and converges to $\theta \approx -5°$. In Fig. 9, the visualization of the subsonic zone in each flow field indicates the variation of the flow structure as $Kn_\infty$ increases. The supersonic jet, which is crucial in Case 1, keeps turning upward and then turns away from the cylinder surface. Consequently, as the flow becomes more rarefied, such a jet impingement upon the cylinder surface gradually fades out.

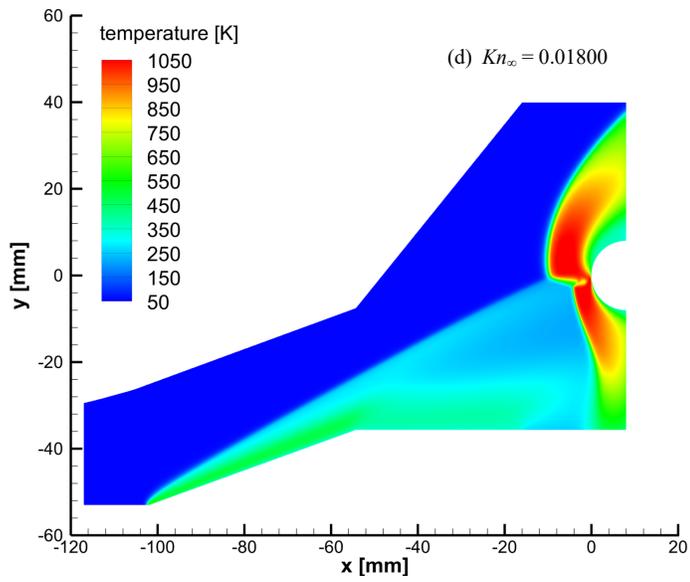
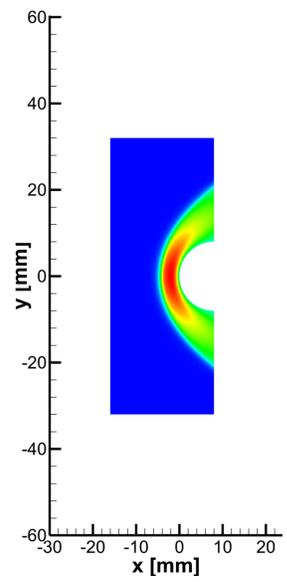

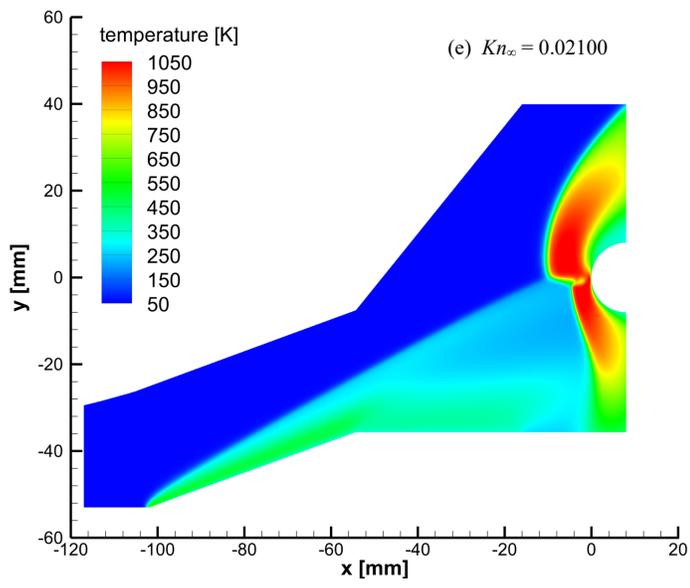
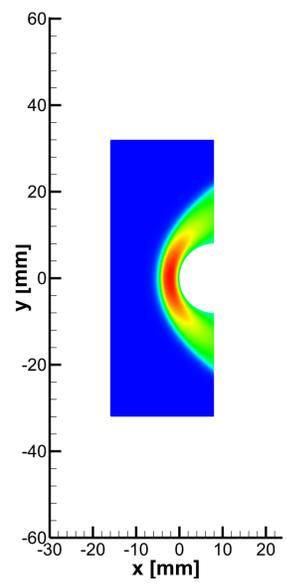

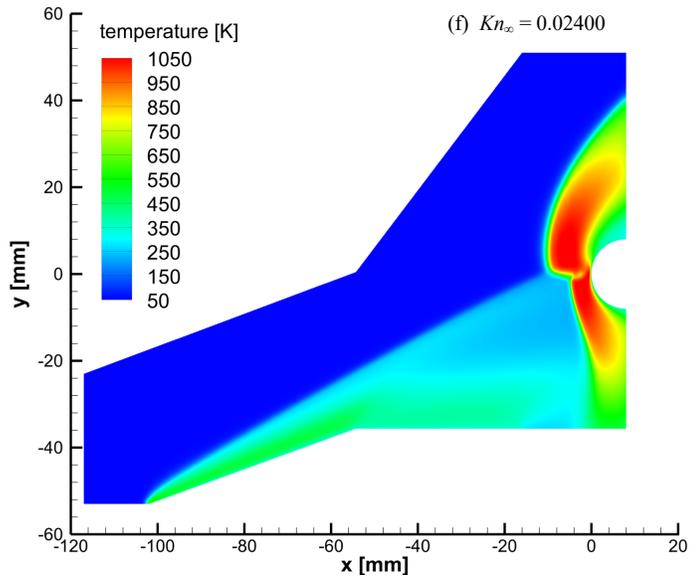
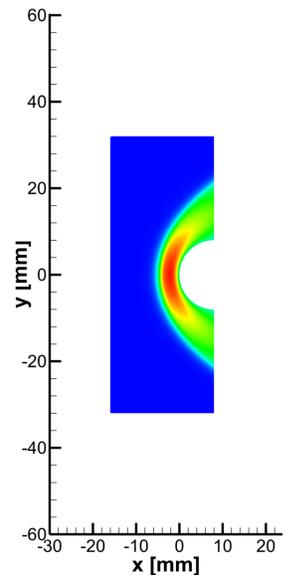

(g) $Kn_\infty = 0.02675$

(h) $Kn_\infty = 0.03290$

(i) $Kn_\infty = 0.04000$

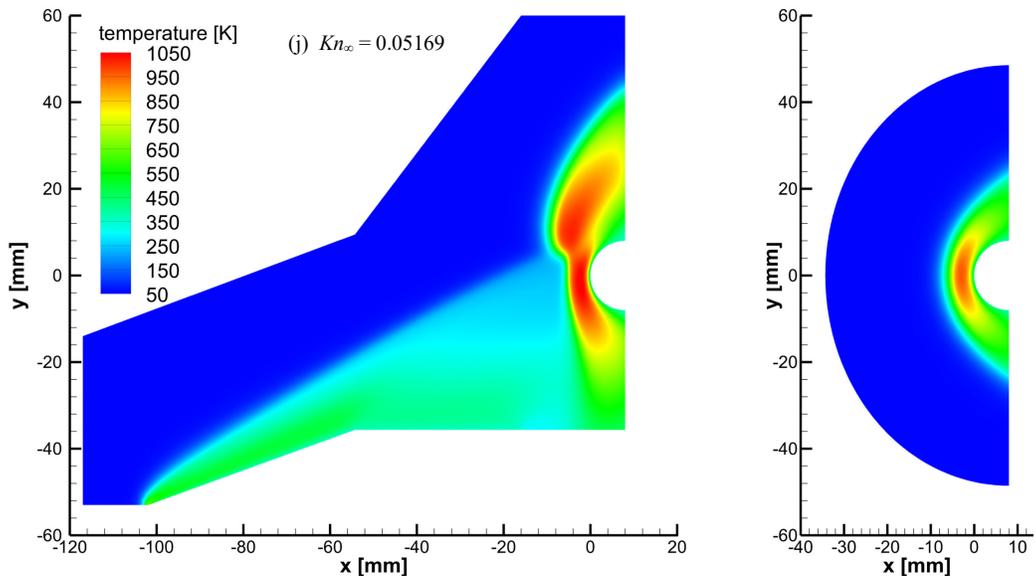
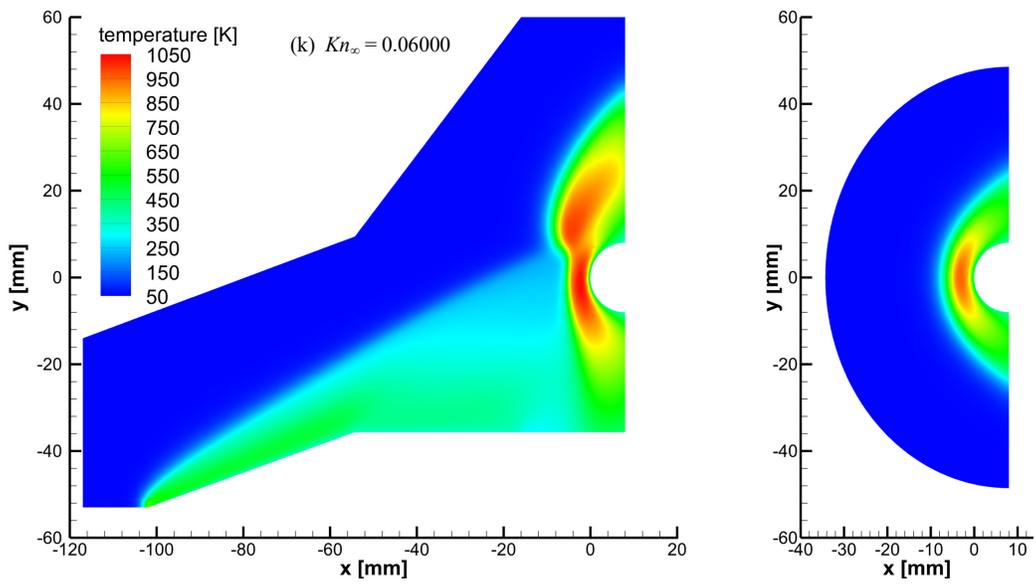
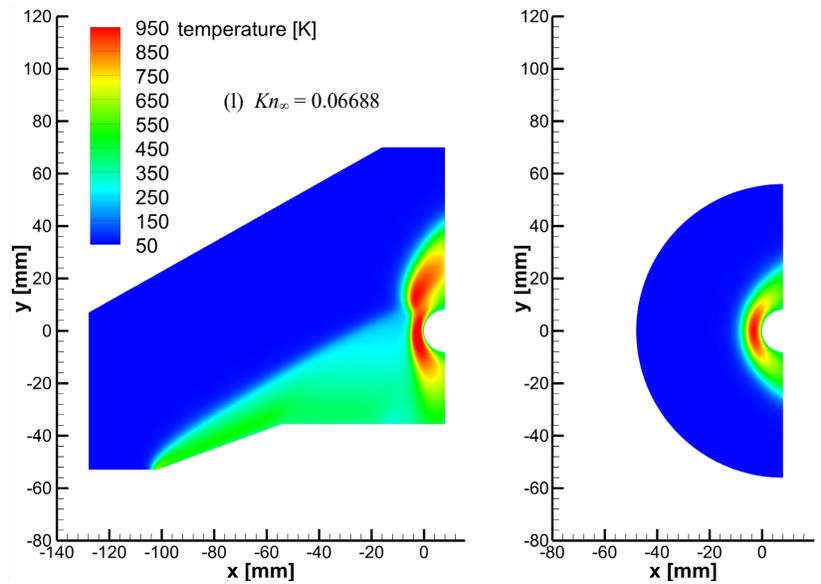

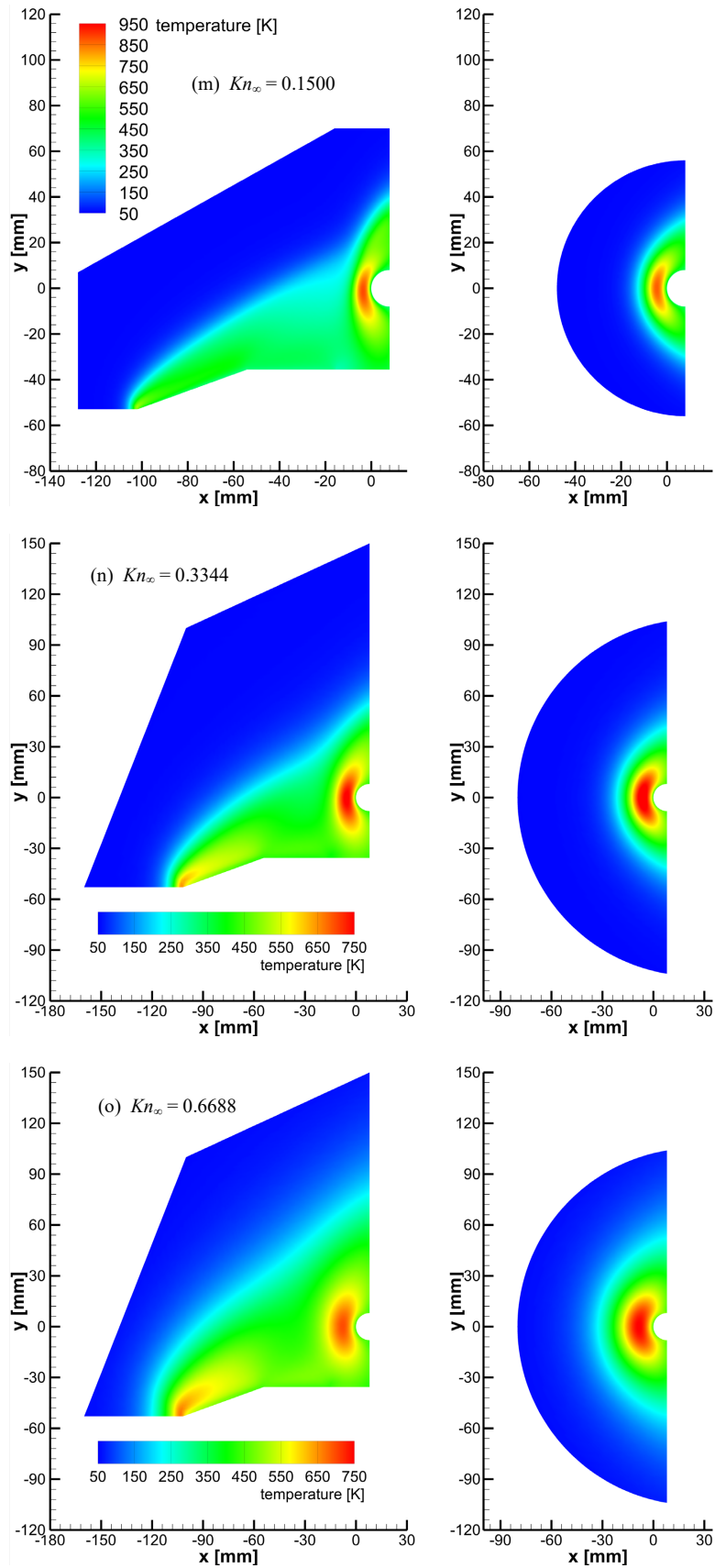

FIG. 7 Temperature fields (with and without shock interaction) for different cases: (a)–(o) correspond to the cases 1–15 in Table II.

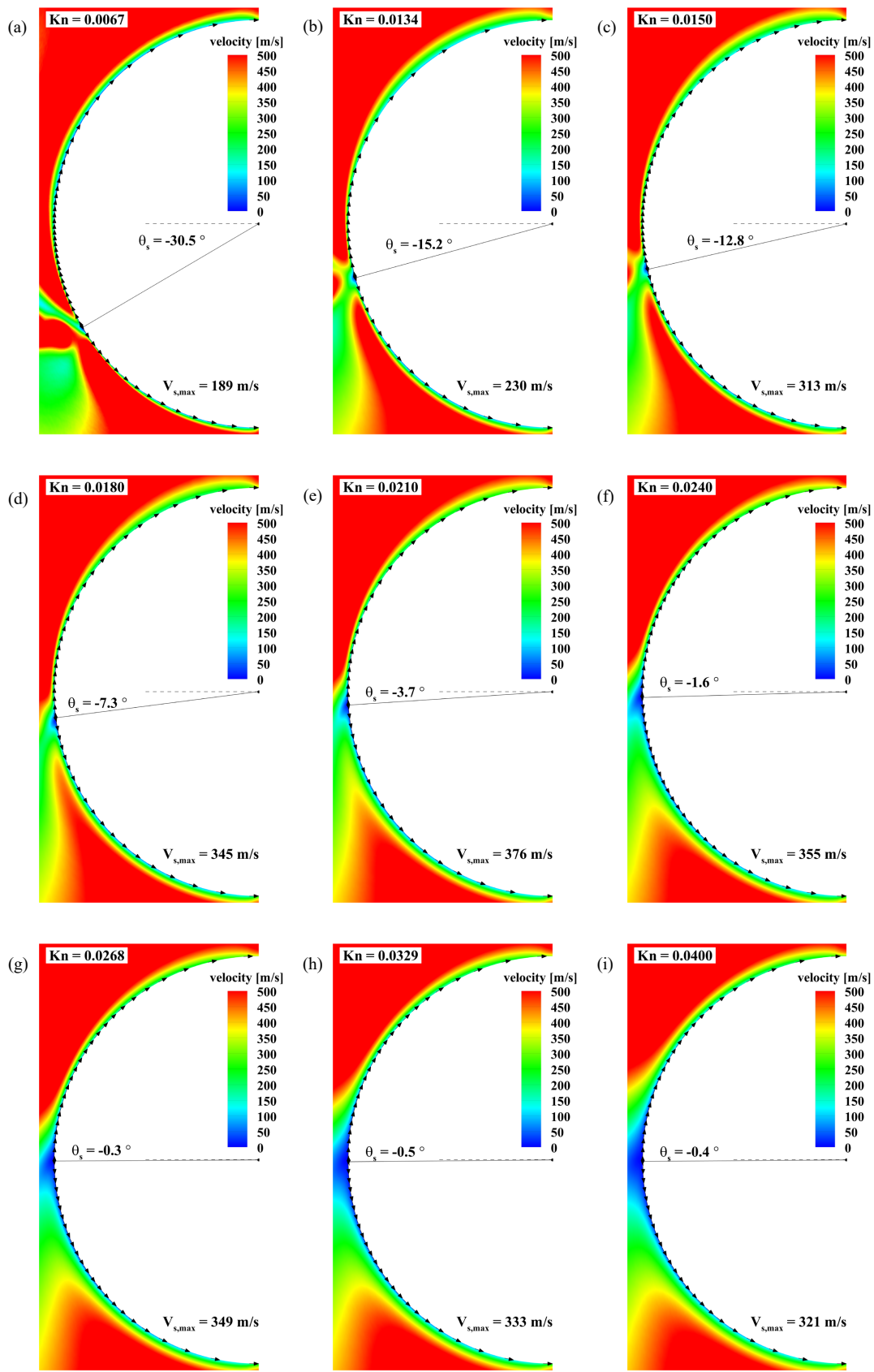

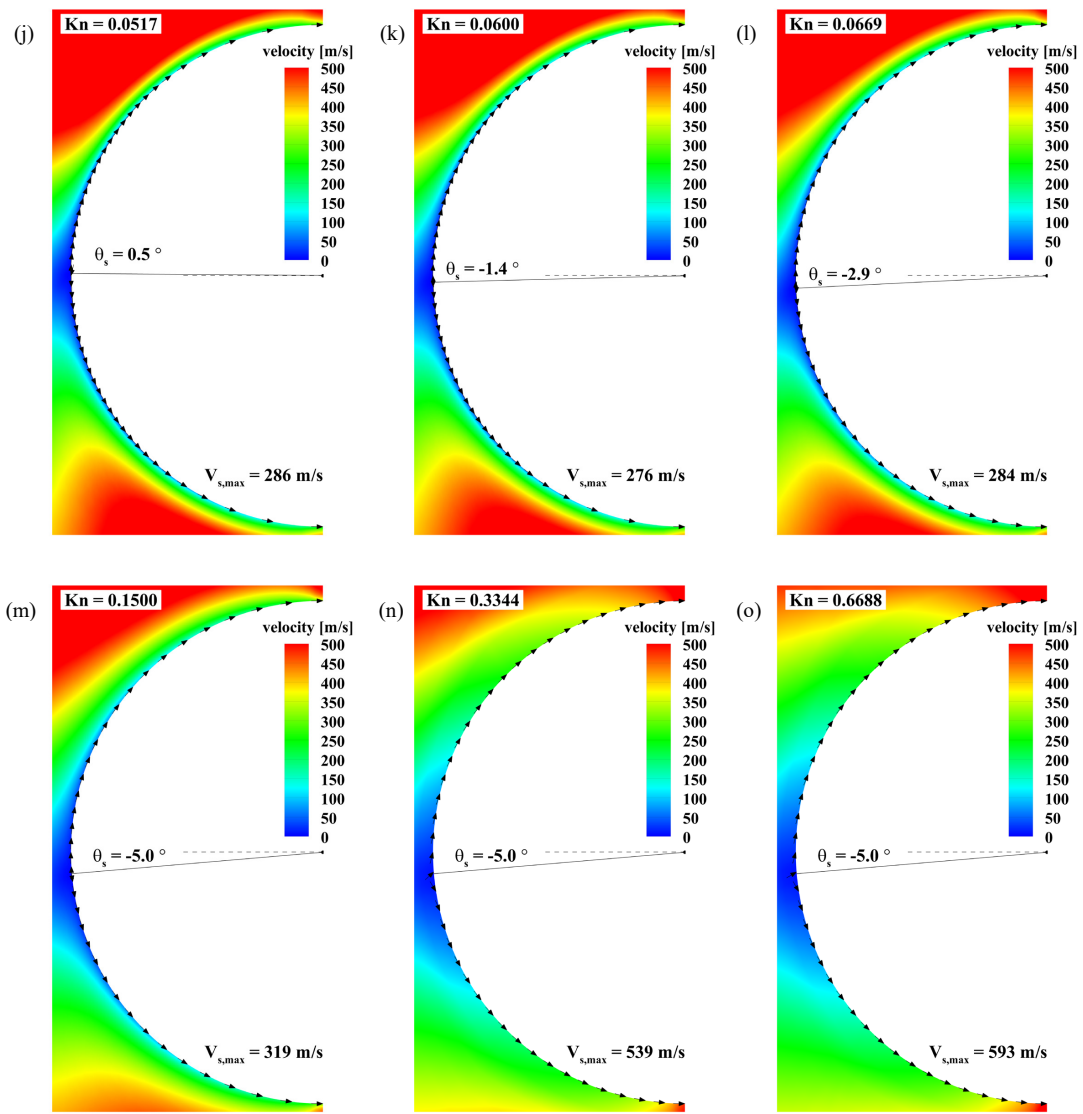

FIG. 8 Velocity fields in the vicinity of the cylinder under shock–shock interactions: (a)–(o) correspond to the cases 1–15 in Table II. The arrows represent the directions of slip velocities on the cylinder surface. The location of stagnation point $\theta_s$ and the maximum magnitude of slip velocity $V_{s,max}$ are marked in each subfigure.

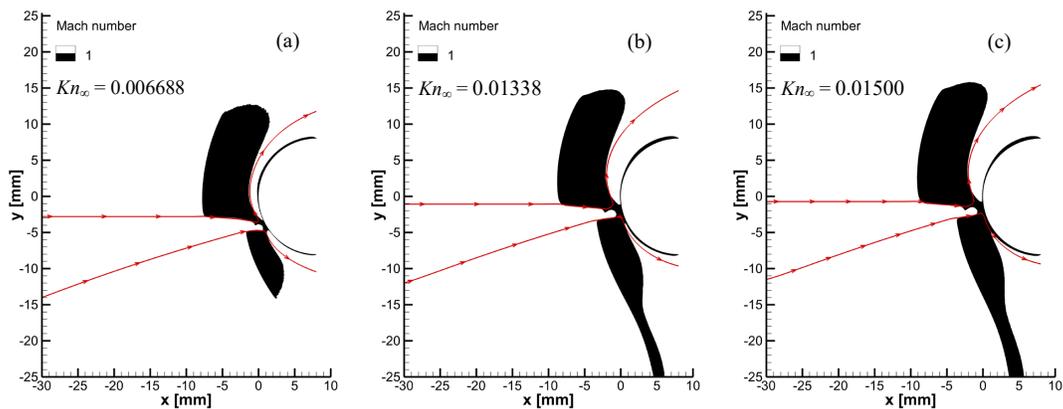

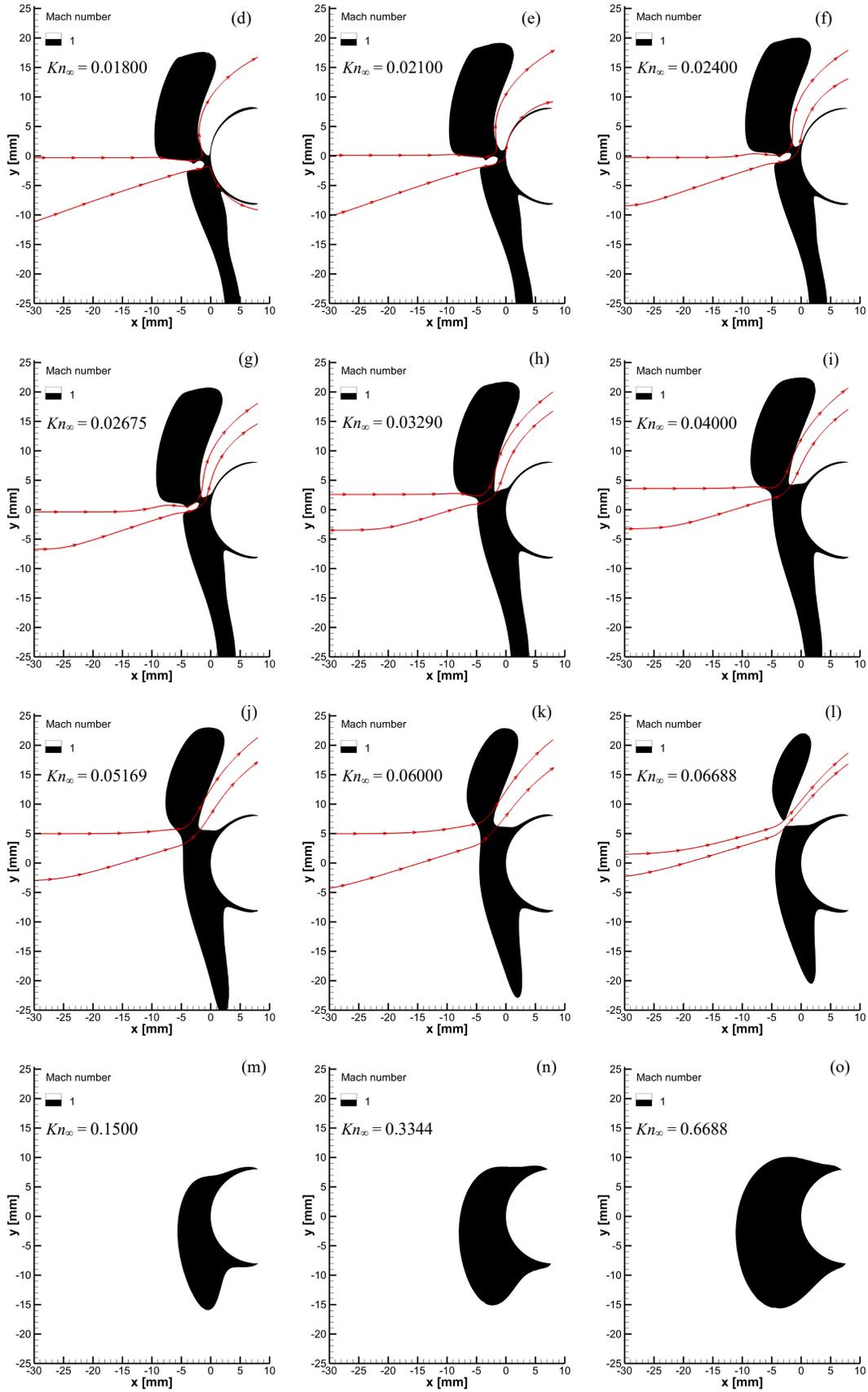

FIG. 9 The subsonic zone and the jet (if exists) formed in the shock–shock interaction: (a)–(o) correspond to the cases 1–15 in Table II.

### *2. Distribution of wall friction*

In Fig. 10, the distributions of the shear stress on the cylinder surface are displayed in ascending order of $Kn_\infty$. In each subfigure of Fig. 10, the results for the 'interaction' and the 'undisturbed cylinder' simulations are both provided.

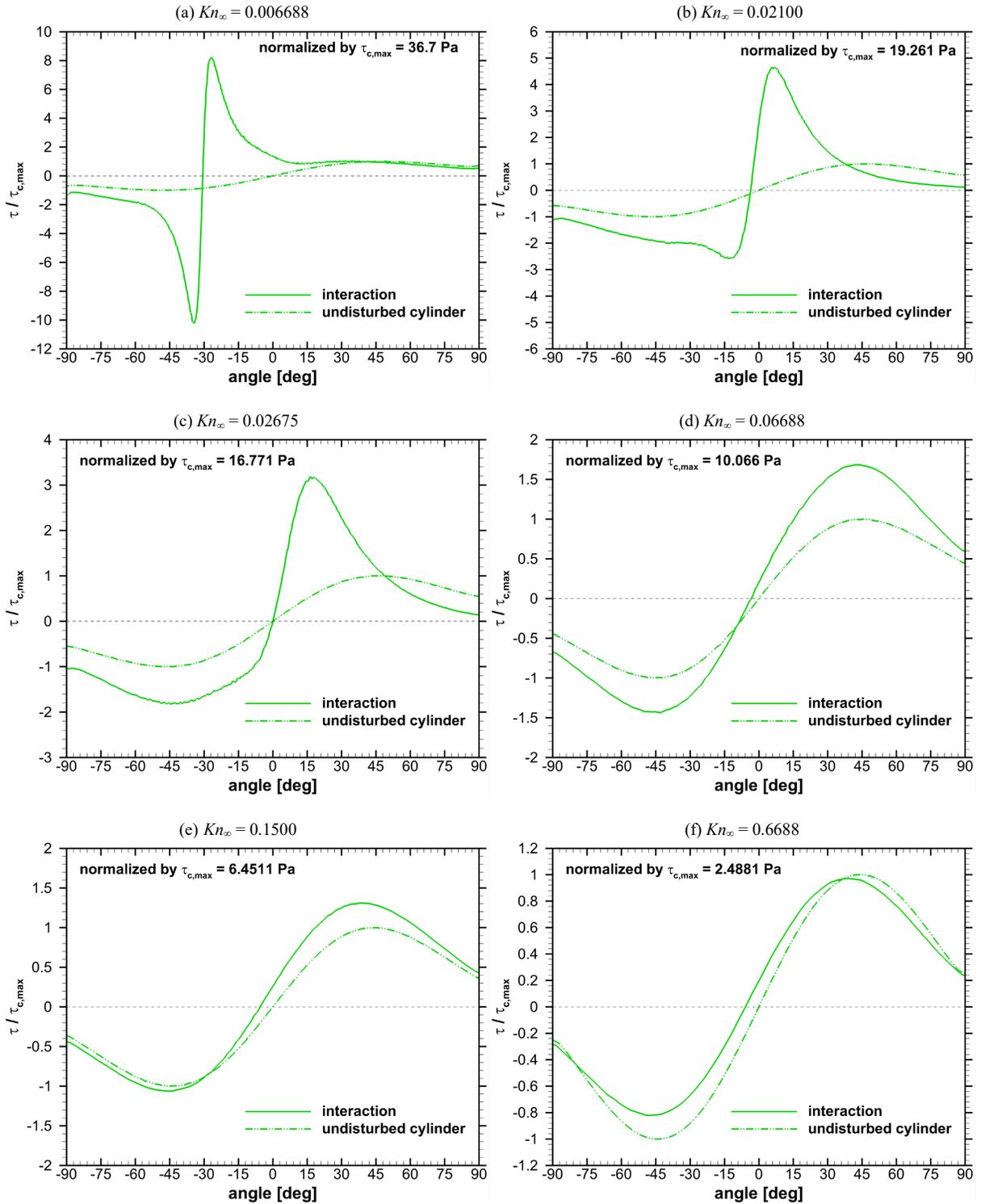

FIG. 10 Distributions of the shear stress on the cylinder surface with and without shock–shock interaction: (a)–(f) correspond to cases of different $Kn_\infty$. The wall shear stress $\tau$ is normalized by the maximum value of $\tau$ in the absence of shock–shock interaction.

In Fig. 10, the positive and negative portions of the friction distribution curve correspond to the different directions of the shear stress. The clockwise circumferential direction is defined as the positive one. Thus, the position of $\tau = 0$ indicates the stagnation point in the flow field. For all considered Knudsen numbers, the friction distributions along the undisturbed cylinder are in a similar form, with the positive and negative peaks at approximately 45° and −45°, respectively. In contrast, the distributions in the shock–shock interaction simulations are significantly affected by the degree of rarefaction.

When the flow is near continuum, as shown in Fig. 10(a), the shock–shock interaction leads to significant amplified surface shear stresses in comparison to $\tau_{c,max}$ for the undisturbed cylinder ($\tau_{min} \approx -10\tau_{c,max}$ and $\tau_{max} \approx 8\tau_{c,max}$). In a very narrow region between the stagnation point and the positive/negative peak (around $\theta = -30°$), the gas stream accelerates rapidly. Figures 10(b)–(f) exhibit the variation of the friction distribution with the increase of $Kn_\infty$. The distribution transits from a highly localized form to a more evenly distributed form. Consequently, the intensities of the peak shear stresses are weakened. Besides, different locations of the positive and negative peaks can be observed as $Kn_\infty$ increases. The angular distance between these two peaks grows from 7.7° to nearly 90°, which indicates that the flow acceleration process from the stagnation point to the downstream region becomes slower when the flow becomes more rarefied.

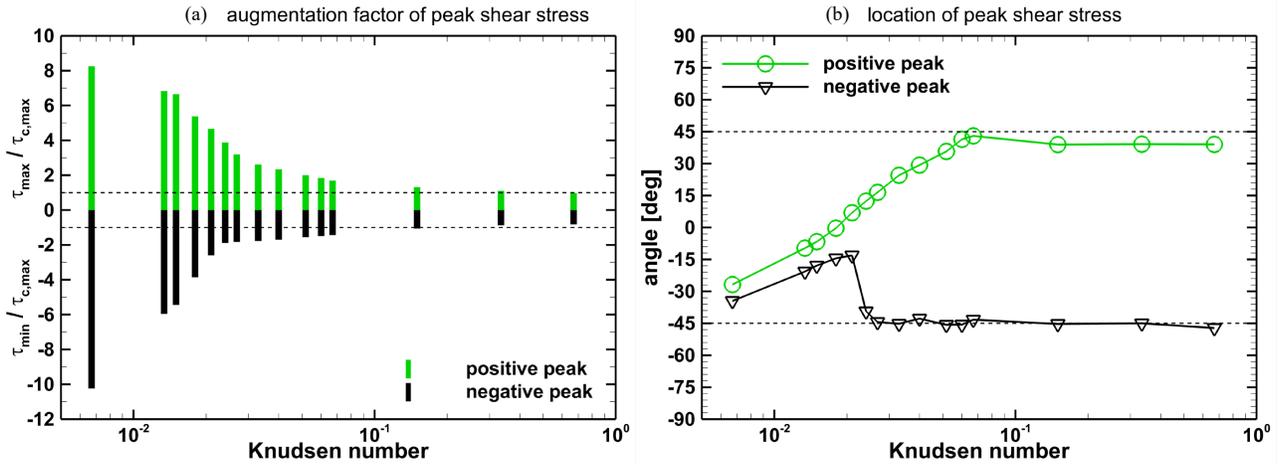

FIG. 11 Augmentation factor and location of the peak shear stress on the cylinder surface in the presence of shock–shock interaction: (a) the augmentation factors of the positive and negative peak values and (b) the locations of the positive and negative peak values.

The augmentation factor is defined as the ratio of the peak shear stress under the shock–shock interaction to $\tau_{c,max}$ on the undisturbed cylinder surface. In Fig. 11, the augmentation factors and the angular positions of the positive and negative peak shear stresses are plotted as functions of $Kn_\infty$, where the dashed lines stand for the values in the absence of shock–shock interaction. The magnitude augmentation and the location shift of the peak shear stresses due to the shock–shock interaction gradually disappear as the flow rarefaction becomes stronger. At $Kn_\infty = 0.6688$, augmentation factors turn out to be less than unity, which means the shock–shock interaction actually reduces the skin friction over the cylinder, as shown in Fig. 10(f).

### *3. Distribution of wall pressure*

In Fig. 12, the distributions of the pressure on the cylinder surface are displayed in ascending order of $Kn_\infty$. In each subfigure of Fig. 12, the results for the 'interaction' and the 'undisturbed cylinder' simulations are both provided.

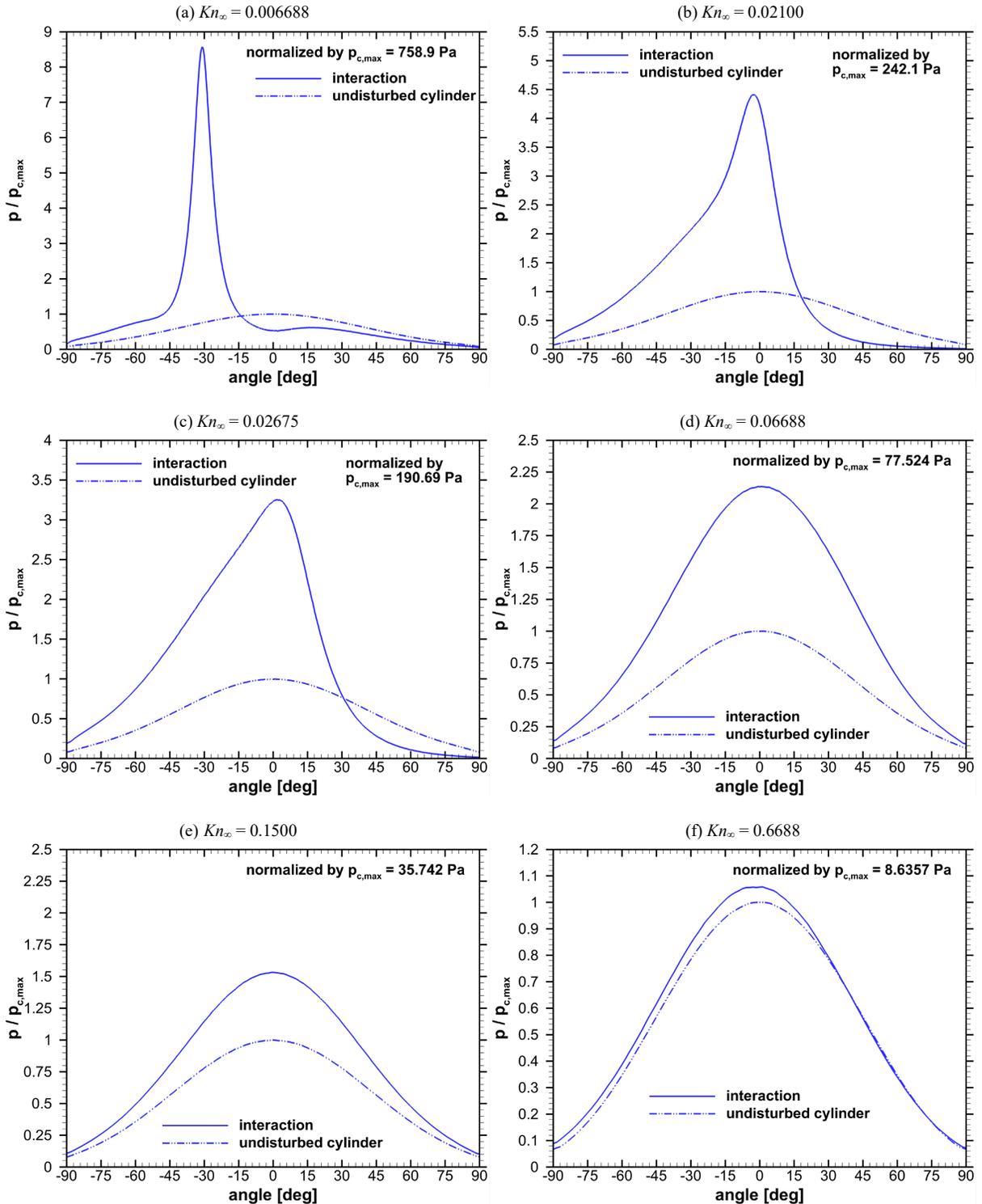

FIG. 12 Distributions of the pressure on the cylinder surface with and without shock–shock interaction: (a)–(f) correspond to cases of different $Kn_\infty$. The wall pressure $p$ is normalized by the maximum value of $p$ in the absence of shock–shock interaction.

Comparison between the solid and dashed lines in Fig. 12 shows that the shock–shock interaction can greatly influence the pressure distribution along the cylinder surface. For all considered Knudsen numbers here, the pressure distributions over the undisturbed cylinder are in a similar form, which is symmetric and has a maximum pressure at $\theta = 0°$. In contrast, the distributions in the shock–shock interaction simulations are significantly affected by the degree of rarefaction.

When the flow is near continuum, as shown in Fig. 12(a), the shock–shock interaction leads to significant amplified surface pressure in comparison to $p_{c,max}$ for the undisturbed cylinder ($p_{max} \approx 8.6\ p_{c,max}$). The peak pressure is located at $\theta = -30°$, and the elevation in wall pressure only occurs in a narrow region around the peak. Figures 12(b)–(f) exhibit the variation of the pressure distribution with the increase of $Kn_\infty$. The distribution transits from a highly localized form to a more evenly distributed form. Consequently, the augmentation effect on the peak pressure is weakened. Besides, different locations of the pressure peaks can be observed as $Kn_\infty$ increases. The angular position of the pressure peak changes from −30° to nearly 0°. In near-continuum regime, the shock–shock interaction greatly shifts the flow stagnation point and thus the location of the peak pressure. However, when the interaction flow becomes more rarefied, the symmetry of the wall pressure distribution is recovered.

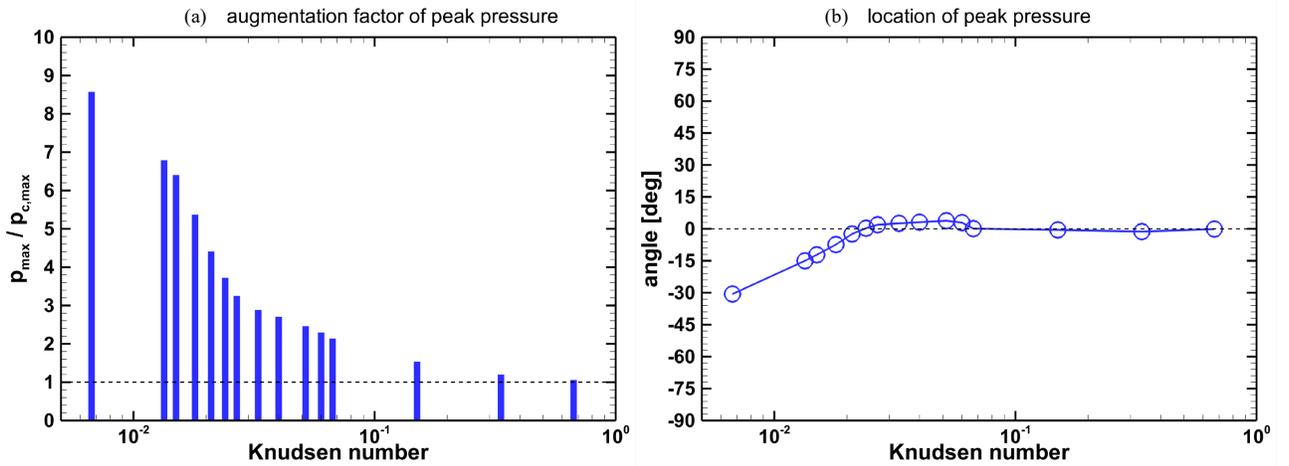

FIG. 13 Augmentation factor and location of the peak pressure on the cylinder surface in the presence of shock–shock interaction: (a) the augmentation factor of the peak value and (b) the location of the peak value.

The augmentation factor is defined as the ratio of the peak pressure under the shock–shock interaction to $p_{c,max}$ on the undisturbed cylinder surface. In Fig. 13, the augmentation factor and the angular position of the peak pressure on the surface are plotted as functions of $Kn_\infty$, where the dashed lines stand for the values in the absence of shock–shock interaction. The magnitude augmentation and the location shift of the peak pressure due to the shock–shock interaction gradually disappear as the flow rarefaction becomes stronger. At $Kn_\infty = 0.6688$, the augmentation factor is only 1.06, which means the shock–shock interaction has negligible influence on the wall pressure, as shown in Fig. 12(f).

## *4. Distribution of wall heat flux*

In Fig. 14, the distributions of the heat flux on the cylinder surface are displayed in ascending order of $Kn_\infty$. In each subfigure of Fig. 14, the results for the 'interaction' and the 'undisturbed cylinder' simulations are both provided.

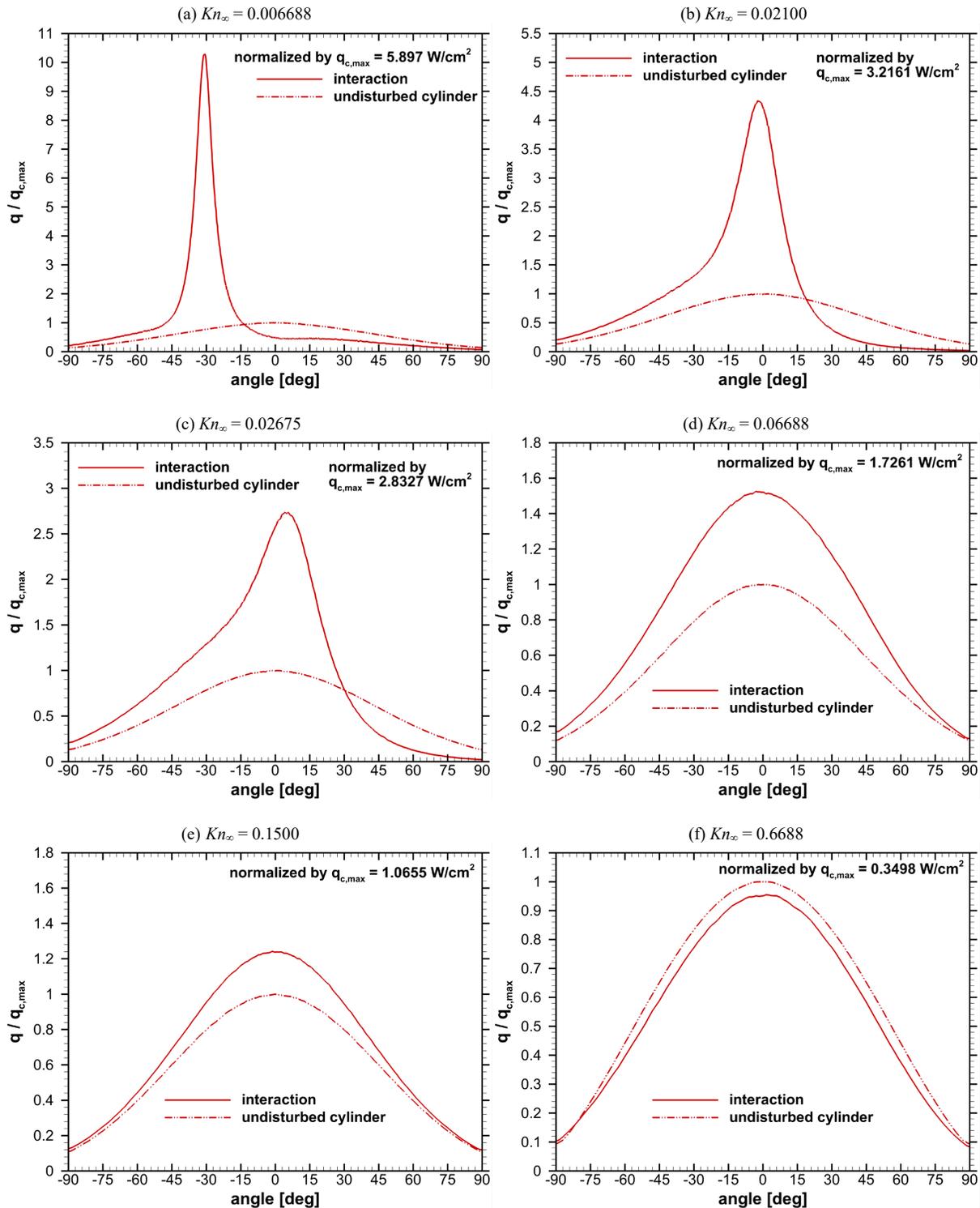

FIG. 14 Distributions of the heat flux on the cylinder surface with and without shock–shock interaction: (a)–(f) correspond to cases of different $Kn_\infty$. The wall heat flux $q$ is normalized by the maximum value of $q$ in the absence of shock–shock interaction.

Comparison between the solid and dashed lines in Fig. 14 shows that the shock–shock interaction can greatly influence the heat flux distribution along the cylinder surface. For all considered Knudsen numbers here, the heat flux distributions over the undisturbed cylinder are in a similar form, which is symmetric and has a maximum pressure at $\theta = 0°$. In contrast, the distributions in the shock–shock interaction simulations are significantly affected by the degree of rarefaction.

When the flow is near continuum, as shown in Fig. 14(a), the shock–shock interaction leads to significant amplified surface heat flux in comparison to $q_{c,max}$ for the undisturbed cylinder ($q_{max} \approx 10.3\, q_{c,max}$). The peak heat flux is located at $\theta = -30°$, and the elevation in heat flux only occurs in a narrow region around the peak. Figures 14(b)–(f) exhibit the variation of the heat flux distribution with the increase of $Kn_\infty$. The distribution transits from a highly localized form to a more evenly distributed form. Consequently, the augmentation effect on the peak heat flux is weakened. Besides, different locations of the heat flux peaks can be observed as $Kn_\infty$ increases. The angular position of the heat flux peak changes from $-30°$ to nearly $0°$. In near-continuum regime, the shock–shock interaction greatly shifts the flow stagnation point and thus the location of the peak heat flux. However, when the interaction flow becomes more rarefied, the symmetry of the wall heat flux distribution is recovered.

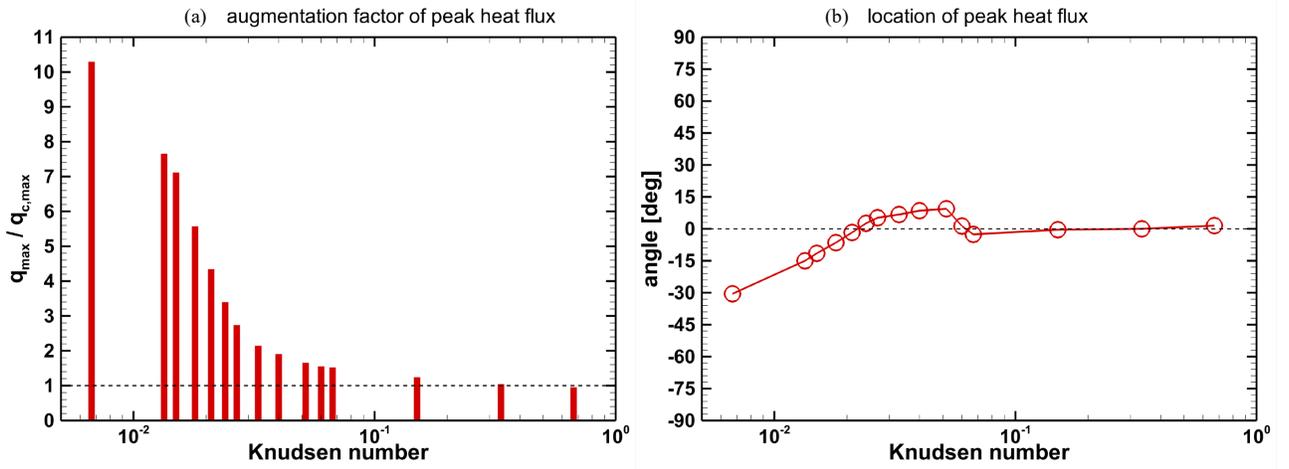

FIG. 15 Augmentation factor and location of the peak heat flux on the cylinder surface in the presence of shock–shock interaction: (a) the augmentation factor of the peak value and (b) the location of the peak value.

The augmentation factor is defined as the ratio of the peak heat flux under the shock–shock interaction to $q_{c,max}$ on the undisturbed cylinder surface. In Fig. 15, the augmentation factor and the angular position of the peak heat flux on the surface are plotted as functions of $Kn_\infty$, where the dashed lines stand for the values in the absence of shock–shock interaction. The magnitude augmentation and the location shift of the peak heat flux due to the shock–shock interaction gradually disappear as the flow rarefaction becomes stronger. At $Kn_\infty = 0.6688$, the augmentation factor turns out to be less than unity, which means the shock–shock interaction actually reduces the aerodynamic heating over the cylinder, as shown in Fig. 14(f).

## C. Discussion

Results in subsection IV. B show that the increasing rarefaction has substantial effects on the shock–shock interaction, in aspects of the flowfield structure as well as the wall surface loads. These effects can be understood by comparing the natures of rarefied and continuum flows. When $Kn_\infty$ is 0.006688 or less, the microscopic process that dominates the gas flow is the collision between gas molecules. The collective effect of these collisions becomes the wave propagation and interference in the context of continuum mechanics. When $Kn_\infty$ grows to the level of 0.6688, the free moves of gas molecules and their reflections on the solid surface greatly influence the flow behavior. Such a rarefied macroscopic flow field is highly diffusive. On this basis, the flow rarefaction can affect the shock–shock interaction through the following mechanisms:

(1) As $Kn_\infty$ increases, the thicknesses of shock waves, shear layers, and wall boundary layers all grow accordingly, and they can no longer be treated as thin layers or discontinuities. At sufficiently large Knudsen numbers, these flowfield structures can merge into a single whole region, where physical fields vary smoothly.

(2) In the shock–shock interaction under continuum condition, the system of waves can effectively deflect the streamlines and concentrate the energy in the flow, especially in Edney type IV interaction. However, as $Kn_\infty$ increases, the wave system gradually loses the abilities of deflection and concentration.

(3) The increasing $Kn_\infty$ alters the boundary condition at solid walls. Strong velocity slip is permitted at the wall boundary and thus affects the position of the stagnation point on the wall surface. Changes of the flow field in the vicinity of the surface can further modify the upstream flow patterns.

(4) As a consequence of (1), the intersection point of the incident shock wave and the bow shock wave keeps moving upward in the transition from Case 1 to Case 15. Indeed, the flow rarefaction alters the shape and the position of the incident shock generated by the wedge. Meanwhile, the standoff distance of the bow shock in front of the cylinder also increases. As a results, the actual intersection point of two shocks keeps moving upward when $Kn_\infty$ grows. The position of the intersection point has great impact on the pattern and strength of the shock–shock interaction. For Case 15, the interaction is so weak that the aerodynamic/aerothermal loads on the cylinder are almost unaffected by the presence of the wedge.

(5) Due to the reasons above, the supersonic jet, which is the key phenomenon in Case 1, is greatly influenced by the increasing rarefaction. As $Kn_\infty$ grows, the jet turns counterclockwise, moves upward, and vanishes gradually. In the transition from Case 1 to Case 15, the supersonic jet is initially terminated by the cylinder. Then, the jet grazes the cylinder. Next, the jet passes over the top of the cylinder without impingement. Finally, the jet disappears in the flow field. The existence of the supersonic jet and its position relative to the cylinder account for the distributions of $\tau$, $p$, and $q$ over the cylinder.

## V. CONCLUSIONS

The interaction between an incident shock generated by a wedge and a bow shock detached from a circular cylinder is numerically investigated. The flow pattern and surface properties under the interaction are compared with the counterparts in the flow past the undisturbed cylinder. With no variation in the freestream Mach number or the geometrical configuration, fifteen different flow conditions with $Kn_\infty$ from 0.006688 to 0.6688 are considered, covering a wide range of flow rarefaction.

Detailed DSMC simulation results demonstrate that the degree of rarefaction has a remarkable impact on the shock–shock interaction phenomenon. The main conclusions can be drawn as follows.

For the condition of $Kn_\infty = 0.006688$, the phenomenon can be clearly identified as the Edney type IV interaction. The increase of $Kn_\infty$ alters the flow pattern continuously, and Edney's classification can be inappropriate to describe the shock–shock interaction in highly rarefied flows.

The effects of shock–shock interaction on the aerodynamic force and heating can be weakened by the increase of flow rarefaction. As the flow becomes more rarefied, the shock–shock interaction will result in smaller augmentation factors and less angular shifts of $\tau_{max}$, $p_{max}$ and $q_{max}$ on the cylinder surface. This variation is nonlinear, and can be divided into two phases. For $0.006688 \leq Kn_\infty \leq 0.06688$, the surface-property distributions keep varying in form. The augmentation factors decline significantly, and the locations of peak values keep moving. However, for $0.06688 \leq Kn_\infty \leq 0.6688$, the surface-property distributions are similar in their forms, and only go through slow declines in the augmentation factors.

At $Kn_\infty = 0.6688$, the distributions of aerodynamic/aerothermal loads with and without the shock–shock interaction are very close. In particular, the augmentation factors for skin friction and heat flux are found to be less than unity, which indicates that the shock–shock interaction actually alleviates the aerodynamic force and heating.

In the transition from the continuum regime to the rarefied regime, the aforementioned observations can be understood by inspecting several flowfield features, including (1) the thickening of shock waves, shear layers and wall boundary layers, (2) the velocity slip on the wall surface, (3) the shift of the intersection point of two shocks, and (4) the orientation change and vanishment of the supersonic jet.

In summary, it is necessary to use gas-kinetic approaches to properly evaluate the impact of shock–shock interactions on the high-altitude aerodynamics characteristics of hypersonic vehicles. Future work will extend this investigation toward the free-molecular regime and figure out how the augmentation factors (e.g., $q_{max}/q_{c,max}$) approach their free-molecular limits.

## ACKNOWLEDGMENTS

This work is supported by the National Natural Science Foundation of China (Grant No. 12302388).

## AUTHOR DECLARATIONS

### Conflict of Interest

The authors have no conflicts to disclose.

### Author Contributions

**Yazhong Jiang:** Conceptualization (lead); Data curation (equal); Formal analysis (lead); Funding acquisition (lead); Investigation (lead); Methodology (lead); Project administration (lead); Resources (equal); Software (lead); Supervision (equal); Validation (equal); Visualization (equal); Writing–original draft (equal); Writing–review & editing (lead). **Tianju Ma:** Data curation (equal); Resources (equal); Supervision (equal); Validation (equal); Visualization (equal); Writing–original draft (equal).

## DATA AVAILABILITY

The data that support the findings of this study are available from the corresponding author upon reasonable request.

## REFERENCES

[1] B. Edney, "Anomalous heat transfer and pressure distributions on blunt bodies at hypersonic speeds in the presence of an impinging shock," FFA Report No. 115, Aeronautical Research Institute of Sweden, 1968.

[2] A. R. Wieting and M. S. Holden, "Experimental shock-wave interference heating on a cylinder at Mach 6 and 8," AIAA J. **27** (11), 1557–1565 (1989).

[3] S. R. Sanderson, H. G. Hornung, and B. Sturtevant, "The influence of non-equilibrium dissociation on the flow produced by shock impingement on a blunt body," J. Fluid Mech. **516**, 1–37 (2004).

[4] Y. Chu and X. Lu, "Characteristics of unsteady type IV shock/shock interaction," Shock Waves **22**, 225–235 (2012).

[5] J. Olejniczak, M. J. Wright, and G. V. Candler, "Numerical study of inviscid shock interactions on double-wedge geometries," J. Fluid Mech. **352**, 1–25 (1997).

[6] D. Wang, Z. Li, Z. Zhang, N. Liu, J. Yang, and X. Lu, "Unsteady shock interactions on V-shaped blunt leading edges," Phys. Fluids **30**, 116104 (2018).

[7] Z. Zhang, Z. Li, R. Huang, and J. Yang, "Experimental investigation of shock oscillations on V-shaped blunt leading edges," Phys. Fluids **31**, 026110 (2019).

[8] E. Zhang, Z. Li, Y. Li, and J. Yang, "Three-dimensional shock interactions and vortices on a V-shaped blunt leading edge," Phys. Fluids **31**, 086102 (2019).

[9] Z. Zhang, Z. Li, and J. Yang, "Transitions of shock interactions on V-shaped blunt leading edges," J. Fluid Mech. **912**, A12 (2021).

[10] D. Wang, G. Han, M. Liu, and Z. Jiang, "Numerical investigation on unsteady interaction of oblique/bow shock during rotation based on non-inertial coordinate system," Phys. Fluids **34**, 121703 (2022).

[11] D. Wang, G. Han, M. Liu, and Z. Jiang, "Numerical investigation of unsteady aerodynamic characteristics induced by the interaction of oblique/bow shock waves during rotation," Phys. Fluids **35**, 086124 (2023).

[12] Y. Wang, Y. Wang, and Z. Jiang, "Unsteady interaction mechanism of transverse stage separation in hypersonic flow for a two-stage-to-orbit vehicle," Phys. Fluids **35**, 056120 (2023).

[13] Y. Wang, Y. Wang, and Z. Jiang, "Experimental and numerical investigation on the unsteady interaction in longitudinal stage separation for parallel-staged two-body configuration," Phys. Fluids **36**, 016116 (2024).

[14] J. Yang, Z. Li, Y. Zhu, Z. Zhai, X. Luo, and X. Lu, "Shock wave propagation and interactions," Advances in Mechanics **46**, 541–587 (2016).

[15] M. S. Ivanov and S. F. Gimelshein, "Computational hypersonic rarefied flows," Annu. Rev. Fluid Mech. **30**, 469–505 (1998).

[16] M. Schouler, Y. Prévereaud, and L. Mieussens, "Survey of flight and numerical data of hypersonic rarefied flows encountered in earth orbit and atmospheric reentry," Prog. Aerosp. Sci. **118**, 100638 (2020).

[17] Z. Wang, L. Bao, and B. Tong, "Rarefaction criterion and non-Fourier heat transfer in hypersonic rarefied flows," Phys. Fluids **22**, 126103 (2010).

[18] Y. Jiang, Z. Gao, C. Jiang, and C.-H. Lee, "Hypersonic aeroheating characteristics of leading edges with different nose radii," J. Thermophys. Heat Transfer **31** (3), 538–548 (2017).

[19] R. Brun, *High Temperature Phenomena in Shock Waves* (Springer, Heidelberg, German, 2012).

[20] H. Alsmeyer, "Density profiles in argon and nitrogen shock waves measured by the absorption of an electron beam," J. Fluid. Mech. **74** (3), 497–613(1976).

[21] G. Pham-Van-Diep, D. Erwin, and E. P. Muntz, "Nonequilibrium molecular motion in a hypersonic shock wave," Science **245** (4918), 624–626 (1989).

[22] H. Hornung, "Regular and Mach reflection of shock waves," Ann. Rev. Fluid Mech. **18**, 33–58 (1986).

[23] M. S. Ivanov, Ye. A. Bondar, D.V. Khotyanovsky, A.N. Kudryavtsev, and G. V. Shoev, "Viscosity effects on weak irregular reflection of shock waves in steady flow," Prog. Aerosp. Sci. **46**, 89–105 (2010).

[24] G. V. Shoev, D.V. Khotyanovsky, Ye. A. Bondar, A.N. Kudryavtsev, and M. S. Ivanov, "Numerical study of triple-shock-wave structure in steady irregular reflection," *27th International Symposium on Rarefied Gas Dynamics*, AIP Conference Proceedings **1333**, 325–330 (2011).

[25] Ye. A. Bondar, G. V. Shoev, A.N. Kudryavtsev, and M. Timokhin, "Nonequilibrium velocity distribution in steady regular shock-wave reflection," *31st International Symposium on Rarefied Gas Dynamics*, AIP Conference Proceedings **2132**, 120005 (2019).

[26] G. V. Shoev, A.N. Kudryavtsev, D.V. Khotyanovsky, and Ye. A. Bondar, "Viscous effects in Mach reflection of shock waves and passage to the inviscid limit," J. Fluid Mech. **959**, A2 (2023).

[27] H. Chen, B. Zhang, and H. Liu, "Non-Rankine–Hugoniot shock zone of Mach reflection in hypersonic rarefied flows," J. Spacecr. Rockets **53** (4), 619–628 (2016).

[28] R. Qiu, Y. Bao, T. Zhou, H. Che, R. Chen, and Y. You, "Study of regular reflection shock waves using a mesoscopic kinetic approach: Curvature pattern and effects of viscosity," Phys. Fluids **32**, 106106 (2020).

[29] Z. Cao, C. White, and K. Kontis, "Numerical investigation of rarefied vortex loop formation due to shock wave diffraction with the use of rorticity," Phys. Fluids **33**, 067112 (2021).

[30] O. Tumuklu, D. A. Levin, and V. Theofilis, "Investigation of unsteady, hypersonic, laminar separated flows over a double cone geometry using a kinetic approach," Phys. Fluids **30**, 046103 (2018).

[31] S. S. Sawant, V. Theofilis, and D. A. Levin, "On the synchronisation of three-dimensional shock layer and laminar separation bubble instabilities in hypersonic flow over a double wedge," J. Fluid Mech. **941**, A7 (2022).

[32] H. Liu, Q. Li, W. Chen, and L. Wu, "On the shock wave boundary layer interaction in slightly rarefied gas," Phys. Fluids **36**, 026115 (2024).

[33] T. Pot, B. Chanet, M. Lefebvre, and P. Bouchardy, "Fundamental study of shock/shock interference in low density flow: Flowfield measurements by DL-CARS," in *21st International Symposium on Rarefied Gas Dynamics*, Marseille, France, (1998).

[34] V. V. Riabov and A. V. Botin, "Shock/shock interference in hypersonic low-density flows near a cylinder," *31st International Symposium on Rarefied Gas Dynamics*, AIP Conference Proceedings **2132**, 100003 (2019).

[35] V. Cardona, R. Joussot, and V. Lago, "Shock/shock interferences in a supersonic rarefied flow: experimental investigation," Exp. Fluids **62**, 135 (2021).

[36] V. Cardona and V. Lago, "Investigation of shock/shock interferences on the aerodynamics of a fragment in the wake of debris in a rarefied regime/at high altitude," J. Fluid Mech. **973**, A26 (2023).

[37] A. B. Carlson and R. G. Wilmoth, "Shock interference prediction using direct simulation Monte Carlo," J. Spacecr. Rockets **29** (6), 780–785 (1992).

[38] J. N. Moss, T. Pot, B. Chanetz, and M. Lefebvre, "DSMC simulation of shock/shock interactions: Emphasis on type IV interactions," in *22nd International Symposium on Shock Waves*, Southampton, UK, (1999).

[39] C. E. Glass, "Numerical simulation of low-density shock-wave interactions," Technical Report, NASA/TM-1999-209358 (1999).

[40] S. Colonia, R. Steijl, and G. Barakos, "Shock interactions in continuum and rarefied conditions employing a novel gas-kinetic scheme," *Shock Wave Interactions*, Springer International Publishing, 107–122 (2018).

[41] C. White and K. Kontis, "The effect of increasing rarefaction on the Edney type IV shock interaction problem," *Shock Wave Interactions*, Springer International Publishing, 299–311 (2018).

[42] M. B. Agir, C. White, and K. Kontis, "The effect of increasing rarefaction on the formation of Edney shock interaction patterns: type-I to type-VI," Shock Waves **32**, 733–751 (2022).

[43] G. A. Bird, *Molecular Gas Dynamics and the Direct Simulation of Gas Flows* (Clarendon Press, Oxford, UK, 1994).

[44] G. A. Bird, *The DSMC Method* (CreateSpace Independent Publishing Platform, San Bernardino, USA, 2013).

[45] I. D. Boyd and T. E. Schwartzentruber, *Nonequilibrium Gas Dynamics and Molecular Simulation* (Cambridge University Press, Cambridge, UK, 2017).

[46] W. Liu, J. Zhang, Y. Jiang, L. Chen, and C.-H. Lee, "DSMC study of hypersonic rarefied flow using the Cercignani–Lampis–Lord model and a molecular-dynamics-based scattering database," Phys. Fluids **33**, 072003 (2021).

[47] Y. Jiang, Y. Ling, and S. Zhang, "Investigation of the inverse Magnus effect on a rotating sphere in hypersonic rarefied flow," Appl. Sci. **14**, 1042 (2024).

[48] R. C. Millikan and D. R. White, "Systematics of vibrational relaxation," J. Chem. Phys. **39**, 3209–3213 (1963).

[49] C. Borgnakke and P. S. Larsen, "Statistical collision model for Monte Carlo simulation of polyatomic gas mixture," J. Comput. Phys. **18**, 405–420 (1975).

[50] *U. S Standard Atmosphere*, (U. S. Government Printing Office, Washington, D. C., USA, 1976).